\documentclass[11pt]{article}
\usepackage{caption}
\usepackage{scalerel}
\usepackage{amsthm,latexsym,amsxtra,graphicx,epstopdf,hyperref,setspace,fix-cm}
\usepackage[normalem]{ulem}
\usepackage{makeidx}
\usepackage{fullpage}
\usepackage{amsmath}
\usepackage{amssymb}
\usepackage{setspace}
\usepackage{bbm}
\usepackage{dsfont}
\usepackage{graphics}
\usepackage{epsfig}
\usepackage[font=footnotesize,labelfont=bf,justification=centerlast,width=.94\textwidth]{caption}
\usepackage{cite}
 \usepackage{multirow}
\usepackage{array,booktabs}
\usepackage{subfigure}
\usepackage{color}
\usepackage[makeroom]{cancel}
\usepackage{pgfplots}
\usetikzlibrary{intersections, pgfplots.fillbetween}
\usepackage{tensor}
\usepackage{simplewick}
\usepackage[most]{tcolorbox}
\usepackage{caption}
\usepackage[skip=0.2cm, indent=0pt]{parskip}

\usepackage{hyperref}

\hypersetup{
    colorlinks,%
    citecolor=blue,
    filecolor=blue,%
    linkcolor=blue,%
    urlcolor=blue
}
\usepackage{float}
\usepackage{slashed}
\usepackage{tikz}
\usetikzlibrary{decorations.pathmorphing}

\graphicspath{{Figures/}}

\def\r{\rightarrow}

\def\J{\mathcal{J}}

\def\b{\beta}
\def\bj{\beta \mathcal{J}}

\newcommand{\bi}{\begin{itemize}}
\newcommand{\ei}{\end{itemize}}
\newcommand{\bea}{\begin{eqnarray}}
\newcommand{\eea}{\end{eqnarray}}
\newcommand{\be}{\begin{equation}}
\newcommand{\ee}{\end{equation}}

\def\XXint#1#2#3{{\setbox0=\hbox{$#1{#2#3}{\int}$}
     \vcenter{\hbox{$#2#3$}}\kern-.5\wd0}}

\numberwithin{equation}{section}
\allowdisplaybreaks  

\usepackage{xcolor}

\usepackage{etoolbox}
\pretocmd{\appendix}{\addtocontents{toc}{\protect\setcounter{tocdepth}{1}}}{}{}

\pgfplotsset{compat=1.18}

\begin{document}
\vspace*{2.5cm}
\begin{center}
\Large \textsc{Exploring the Infrared Landscape of the SYK Model}\\ \vspace*{0.5cm}
\vspace*{1cm}
\end{center}

\begin{center}
Weam Abou Hamdan and Dami\'an A. Galante
\\
\vspace{0.5cm}
\end{center}

\begin{center}
{
\normalsize
Department of Mathematics, King's College London, Strand, London WC2R 2LS, UK

}
\end{center}

\begin{center}
{\textsf{{
\nolinkurl{weam.abou\_hamdan@kcl.ac.uk},  \nolinkurl{damian.galante@kcl.ac.uk}}} } 
\end{center}

\vspace*{0.5cm}

\vspace*{1.5cm}
\begin{abstract}

\noindent We analyse a class of SYK models whose Hamiltonian is the sum of two SYK Hamiltonians with different numbers of fermions $q, \tilde q$ in each interaction. We consider both Euclidean and Lorentzian probes of the quantum system in the large $N$ limit. In the strong coupling phase, the entropy provides a diagnostic of the thermal renormalisation group flow. Under certain conditions, two parametrically separated regimes of near-conformal behaviour emerge. The first reproduces the standard linear-in-temperature scaling characteristic of the single SYK model. The system then flows to another near-fixed point whose entropy scaling depends on the ratio $n = q/\tilde q$. For $n<3/2$, the entropy exhibits anomalous, stronger-than-linear scaling in temperature. At $n=3/2$, there is an additional logarithmic enhancement. Using conformal perturbation theory, we argue that in the infrared regime of the SYK model, there may exist disordered conformal operators with dimensions $1 < \Delta \leq 3/2$. In Lorentzian signature, we study the out-of-time-ordered correlator and show that these deformed theories exhibit near-maximal chaos in both regimes (when they exist). We comment on the relation between the anomalous scalings found here and those observed in certain near-extremal black holes in two and higher dimensions.

\vspace{1cm}

\bigskip

\end{abstract}

\newpage

\setcounter{tocdepth}{2}

\tableofcontents

\section{Introduction} \label{sec:intro}

Near-extremal black holes provide an ideal setting for exploring quantum properties of spacetime. Their near-horizon geometry features an infinite throat region which, in $d$ spacetime dimensions, quite universally takes the form of AdS$_2 \times S^{d-2}$. This allows for explorations at both the macro and microscopic levels \cite{Maldacena:1998uz}.

A natural question is whether further structure can exist beyond the AdS$_2$ throat and, if so, how to probe it. Examples of such structure have been known for a long time, including the solutions by Majumdar \cite{PhysRev.72.390} and Papapetrou \cite{papapetrou} from 1947, in which the AdS$_2$ throat is fragmented into a collection of new AdS$_2$ throats residing in the interior of the original one. Some modern reincarnations include, for instance, \cite{Faulkner:2009wj, Anninos:2011vn, Anninos:2013nra, Bena:2018bbd,Mirfendereski:2018tob}.

A closely related question concerns the smoothness of horizons in the (near-)extremal limit. The Majumdar–Papapetrou geometries already provide explicit examples where smoothness fails. More recently, it has been shown that, for asymptotically AdS, charged black holes in four and five dimensions, \textit{generic} 
non-spherical perturbations lead to horizons that are non-smooth and, in some cases, even singular \cite{Horowitz:2022mly, Horowitz:2022leb}. At the linearised level, consider a metric perturbation $h$ that scales as $h \sim \rho^{\gamma/2}$, where $\rho$ is the Gaussian null coordinate such that the horizon resides at $\rho=0$. Notice that these perturbations are singular for $\gamma<0$ and non-smooth for $0<\gamma \neq 2<4$. Although these deformations keep all curvature invariants finite when $\gamma>0$
, the linearised Weyl and Ricci tensors behave as
\begin{equation}
    \label{null divergences}
    \delta C_{\rho a \rho b} \sim \frac{\gamma}{2}\left(\frac{\gamma}{2} -1\right)\rho^{\tfrac{\gamma}{2} - 2} \, \, , \, \, \delta R_{\rho \rho} \sim \frac{\gamma}{2} \left(\frac{\gamma}{2} - 1\right)\rho^{\tfrac{\gamma}{2} - 2} \, ,
\end{equation}
which clearly diverge as $\rho \to 0$ when $0<\gamma \neq 2 <4$. 
This behaviour was later generalised to a variety of setups \cite{Horowitz:2023xyl, Horowitz:2024dch, Horowitz:2024kcx, Horowitz:2025ayc}, while also studied from an effective field theory perspective \cite{Chen:2024sgx}.

In relation to holography, a natural probe for these features is the thermodynamic behaviour that stems from the Euclidean gravitational path integral \cite{Gibbons:1976ue}. In fact, in the saddle-point approximation where the gravitational constant $G_N$ goes to zero, the low-temperature entropy of near-extremal black holes scales as
\begin{equation}
    \label{general form of entropy}
    \frac{S}{N}=S_0+\frac{S_1}{\beta^{\gamma}}+\cdots \,,
\end{equation}
where $S_0$ is known as the zero-temperature entropy, $\beta$ is the inverse temperature, and $N$ is some large parameter associated with the inverse of $G_N$.\footnote{For instance, the precise match between the infrared behaviour of the SYK model and JT gravity yields,  in the large $q$ limit, $\tfrac{2\pi^3 N}{q^2} = G_{N}^{-1}$.}
In the mostly studied, spherically symmetric, near-AdS$_2$ case, we have $\gamma = 1$, so that the entropy grows linearly with temperature. Given that the entropy is dimensionless, $S_1$ needs to have, in the linear case, units of inverse temperature. It is now understood that those units come from the soft breaking of the conformal symmetry in the infrared, which is given by a Schwarzian action \cite{Maldacena:2016upp}. This Schwarzian action is also responsible for a one-loop exact logarithmic contribution, $S =  -\tfrac{3}{2} \log \beta$, which dominates the thermodynamics at very low temperatures $\beta \log\beta \gg N$ (in units of the black hole energy gap) and, in the absence of supersymmetry, makes the density of states go to zero at absolute-zero temperature \cite{Stanford:2017thb , Iliesiu:2020qvm}. 

However, this does not seem to be the generic case for near-extremal black holes. For some of the deformed near-extremal black holes found in \cite{Horowitz:2022mly}, the entropy as a function of temperature was found to have anomalous scaling, including
\begin{equation}
        0<\gamma< 1 \,.
\end{equation}
In these cases, the entropy decays at a stronger-than-linear rate as the temperature goes to zero. In the particular case of $\gamma \to 1$, there is a logarithmic enhancement such that $S-S_0 \sim \tfrac{\log \beta}{\beta}$. Assuming the result extrapolates to $\gamma<0$, the entropy naively diverges in the zero-temperature limit. These results are all valid semiclassically, to leading order in the saddle-point approximation, and seem to compete with potential one-loop contributions. It would be interesting to see how the finite-$N$ effects in these deformed cases change with respect to the spherically symmetric one.

From the microscopic perspective, the Sachdev-Ye-Kitaev (SYK) model \cite{Sachdev:1992fk , kitaev_vid, Maldacena:2016hyu} encodes many of the near-extremal physics described so far.\footnote{In addition, the Jackiw-Teitelboim (JT) gravity partition function which arises in the near-extremal limit can be recast in terms of a matrix integral \cite{Saad:2019lba}. Studies of deformations in this matrix model context include \cite{Maxfield:2020ale, Witten:2020wvy, Turiaci:2020fjj, Eberhardt:2023rzz,Kruthoff:2024gxc}.}  This disordered model of $N$ fermions subject to all-to-all interactions has an entropy which, in the large $N$ limit and at low energies, takes the form of \eqref{general form of entropy} with $\gamma=1$. It was further shown that, in this parametric regime, this model exhibits an emergent conformal symmetry which is softly broken by a Schwarzian action, in analogy to the gravitational case \cite{Maldacena:2016hyu}.\footnote{Note, however, that as opposed to other known examples of holography, in the case of SYK, there is no limit in which the putative dual bulk fields would decouple from the gravitational description. Their masses always remain of order one \cite{Gross:2017hcz}.} The partition function of the SYK model has, after averaging over disorder, one coupling constant $\mathcal{J}$ with units of energy, so that at finite temperature there is only one dimensionless parameter $\bj$.

Given the analogy to near-extremal black holes, one could ask whether there exist deformations of the SYK model that can affect the low-temperature behaviour of the model and, in particular, the thermodynamic entropy. From the quantum theory perspective, this amounts to deforming the SYK Hamiltonian with a relevant deformation which, at finite temperatures, triggers a thermal renormalisation group (RG) flow \cite{CASTRONETO1993525,Zabzine:1997gh,Appelquist:1999hr, Delacretaz:2021ufg}, in which $\bj$ can be taken to be the energy scale. In the case of SYK, there is an additional complication since `simple' single-trace operators in the model become irrelevant at low energies \cite{Gross:2017hcz}. This can be overcome by considering the following deformed SYK Hamiltonian,
\begin{equation}
    \label{deformed Hamiltonian}
    H = H_q + s H_{\tilde q} \,.
\end{equation}
Here $H_x$ denotes a single SYK Hamiltonian with fermion interactions in groups of $x$, and $s$ is a real dimensionless coupling that measures the ratio between the SYK coupling of each separate Hamiltonian. At large $N$, when $\tilde q<q$ and $s$ is small, it was indeed shown that in the strong coupling regime $\bj\gg1$, there exist two regions that exhibit near-conformal physics \cite{Anninos:2022qgy}.

Given that the $q=2$ model is integrable, Hamiltonians like \eqref{deformed Hamiltonian} with $\tilde q=2$ were first studied as models for chaotic-to-integrable transitions \cite{Garcia-Garcia:2017bkg, Kim:2020mho, Garcia-Garcia:2020dzm, Lunkin:2020tbq, Nandy:2022hcm,Menzler:2024atb,Louw:2023lpq, Lau:2023pot, Lau:2025dgd}. More generally, analytical control at large $N$ can be gained by also taking the large $q,\tilde q$ limit, while keeping $n\equiv q/\tilde q$ fixed. The case of $n=2$ was analysed in \cite{Jiang:2019pam, Anninos:2020cwo, Khveshchenko:2023upm, Khveshchenko:2022gcd}. In \cite{Anninos:2022qgy}, a systematic study of these deformed SYK models was initiated away from the $n=2$ case, while dynamical properties of these models were studied in \cite{Chapman:2024pdw}. The main conclusion can be summarised as follows. 

Consider the deformed SYK model at large $N$. When $s \ll 1$, the entropy of the model has two parametric scaling regimes at $\bj \gg 1$. At low but not-so-low temperatures,\footnote{``Not-so-low'' is precisely defined in section \ref{sec: intermedIR}, see \eqref{intermed inequality}. Recall that $s$ here gives an extra dimensionless variable to parametrically separate regimes.} the entropy is dominated by the entropy of the single SYK model with $q$-fermions and does not depend on the perturbation parameter $s$. We call this regime the intermediate infrared (IR). It is the first correction away from the intermediate IR that depends on $s$ and leads the RG flow towards the deep IR regime, in which the temperature is lower than any other scale, and a new linear-in-temperature regime emerges. Thus, at low temperatures we have
\begin{equation}
    \begin{cases}
        \frac{S}{N} =S_0^{\text{free}}-\frac{\pi^2}{4q^2}+\frac{\pi^2}{q^2}\frac{1}{\beta \mathcal{J}}-\frac{2\beta \mathcal{J}s^2}{q^2}+\mathcal{O}\!\left(\!\frac{1}{(\beta \mathcal{J})^2}\!\right) \,, &\text{Intermediate IR}. \\
        \frac{S}{N} =S_0^{\text{free}}-\frac{\pi^2}{4\tilde{q}^2}+\frac{\sqrt{1+4s^2}}{2s^2}\frac{\pi^2}{\tilde{q}^2}\frac{1}{\beta \mathcal{J}}+\mathcal{O}\!\left(\! \frac{1}{(\beta \mathcal{J})^2} \!\right) \,, &\text{Deep IR}.
    \end{cases}
    \label{eq: intro RG}
\end{equation}
In this paper, we continue the exploration of these deformed SYK models in both Euclidean and Lorentzian signatures. Particular emphasis is placed on cases where $n$ is close to one. Consistent with \eqref{eq: intro RG}, we find that away from the linear-in-temperature intermediate regime, the entropy scales with a coefficient $s^2(\bj)^{2-\frac{2}{n}}$, which resembles the cases of $\gamma = \frac{2}{n}-2<0$ in \eqref{general form of entropy}. Although the entropy may seem divergent for all $n>1$, the flow continues at low temperatures to another smooth parametric regime, which is the deep IR. There, we find that for $n>3/2$, the entropy takes the form of \eqref{general form of entropy} with $\gamma=1$, so there is another region of linear-in-temperature entropy in the deep IR. In this case, the coefficient $S_1$ does \textit{not} take the value of a single SYK with coupling $s\J$. 

However, for $n<3/2$, we find that the deep IR entropy scales in an anomalous way, with
\begin{equation}
 \gamma (n<3/2) = 2n-2 < 1\,, 
 \label{eq: anomalous}
\end{equation}
providing examples of SYK models where the linear-in-temperature entropy does not dominate at low temperatures.\footnote{Other SYK models which exhibit curious infrared behaviour are the supersymmetric SYK models in \cite{Anninos:2016szt,Biggs:2023mfn}. In those cases, however, the leading finite-temperature contribution to the entropy has $\gamma>1$.} Similar physics was previously obtained in a theory with two coupled SYK models \cite{Milekhin:2021cou}. The anomalous scaling \eqref{eq: anomalous} is compatible with the result of conformal perturbation theory for an operator of dimension $1<\Delta=n<3/2$ away from the deep IR near-fixed point, and resembles the one for the deformed near-extremal black holes of \cite{Horowitz:2022mly}. 

We provide both analytical and numerical evidence for this scaling both at infinite and finite $q, \tilde q$. For the special case of $n=3/2$, there is a logarithmic enhancement such that the entropy scales, at the lowest temperatures, as $\tfrac{\log \bj}{\bj}$. Furthermore, we find that the zero-temperature entropy at large $N$ is independent of $s$ (clarifying certain aspects in \cite{Anninos:2022qgy}), and that the deformed SYK models are maximally chaotic in the deep IR, despite the anomalous scaling of the entropy. Even at strong coupling, these models are under such computational control that it seems feasible to access $1/N$-corrections in the future.

The remainder of the paper is structured as follows. In section \ref{sec: deformedSYK}, we provide a short review of the deformed SYK models. In section \ref{sec: DeepIR}, we study the deep infrared thermodynamics, showing universal results for the zero-temperature entropy and anomalous scaling for the leading finite-temperature correction. In section \ref{sec: intermedIR}, we provide analytical and numerical evidence for the existence of an intermediate infrared regime and analyse its main properties. In section \ref{sec: OTOCs}, we provide a real-time probe of these deformed SYK models by studying out-of-time-correlation functions. 
We end with some comments in relation to the holographic duals of the deformed SYK models in section \ref{sec:outlook}. Appendices \ref{sec: appendix A}-\ref{app: multiple} provide further technical details of the computations in the main text.



\section{The deformed SYK model} \label{sec: deformedSYK}

The Sachdev-Ye-Kitaev (SYK) model is a (0+1)-dimensional quantum theory of $N$ Majorana fermions furnishing a finite-dimensional representation of the Clifford algebra
\begin{equation}
\label{Clifford Algebra}
    \{\psi_{i},\psi_j\}=\delta_{ij} \, ,
\end{equation}
for $i,j=1, \ldots ,N$.
Given an even integer $q \geq 2$, these fermions interact randomly via the Hamiltonian
\begin{equation}
\label{SYK Hamiltonian}
    H_q=i^{q/2} \sum_{1 \leq i_1< \ldots <i_q \leq N} J_{i_1 \ldots  i_q}\psi_{i_1} \cdots \psi_{i_q} \, ,
\end{equation}
where the coupling constants $J_{i_1 \ldots i_q}$ independently follow a Gaussian distribution centred at zero with width
\begin{equation}
\label{Gaussian width}
    \langle J_{i_1 \ldots i_q}^2\rangle = \frac{2^{q-1} (q-1)!}{q} \frac{\mathcal{J}^2}{N^{q-1}} \, ,
\end{equation}
specified by an energy scale $\mathcal{J}$. The model is usually considered as an ensemble of Hamiltonians where, in order to compute observables of the theory, we are instructed to sum over all possible realisations of the couplings. 

The large $N$ limit of this theory exhibits time-reparametrization invariance in the strict infrared limit, where the fermions act as primary operators of conformal weight $\Delta=1/q$. However, 
a finite $\J$ explicitly breaks the conformal symmetry, which is, in fact, also spontaneously broken. 
The leading contribution to the action of this near-conformal theory is then given by 
the Schwarzian action. Different reviews of this model include \cite{Sarosi:2017ykf , Rosenhaus:2018dtp, Trunin:2020vwy}.

Throughout this paper, we study a variant of this theory, which we refer to as the deformed SYK model, with a Hamiltonian given by
\begin{equation}
\label{Deformed Hamiltonian}
    H=H_q+ s  H_{\tilde{q}} \quad , \quad s\in \mathbb{R} .
\end{equation}
This deformed Hamiltonian consists of a sum of two standard SYK Hamiltonians, each with a different number of fermions in the interaction. The couplings in both Hamiltonians are randomly chosen from Gaussian distributions with the same $\J$. The dimensionless coupling $|s|$ then measures the relative energy scale between $H_q$ and $H_{\tilde q}$. As we will see shortly, when $\tilde q < q$ and $s$ is small, we can think of $H_{\tilde q}$ as a disordered conformal operator that triggers a renormalisation group flow away from the near-conformal fixed point of the single SYK model.

As with the single SYK model, these deformed models can be studied at large $N$ after averaging over the disorder. To see this, it is convenient to describe the theory in terms of the Euclidean time-ordered propagator for the fermions,
\begin{equation}
\label{Propagator}
    G(\tau_1,\tau_2)=\frac{1}{N} \sum_{i=1}^{N} \langle T\psi_{i}(\tau_1)\psi_i(\tau_2)\rangle_\beta \, ,
\end{equation}
where $\langle \cdot \rangle_{\beta}$ is the expectation value in the thermal state with inverse temperature $\beta$, and $\tau$ is the Euclidean time with periodic identification $\tau \sim \tau+\beta$. In the large $N$ limit, the leading contributions to this quantity arise from melonic Feynman diagrams, so that together with the self-energy bilocal field $\Sigma(\tau_1,\tau_2)$, the two-point function satisfies (deformed) Schwinger-Dyson equations,
\begin{equation}
\label{Schwinger-Dyson}
    \begin{cases}
        G^{-1}(\tau_1,\tau_2) = \delta(\tau_1-\tau_2)\partial_{\tau_2}-\Sigma(\tau_1,\tau_2) \, , \\
        \Sigma(\tau_1,\tau_2) = \mathcal{J}^2 \left( \frac{2^{q-1}}{q}  G(\tau_1,\tau_2)^{q-1}+s^2\frac{2^{\tilde{q}-1}}{\tilde{q}}  G(\tau_1,\tau_2)^{\tilde{q}-1} \right) \, .
    \end{cases}
\end{equation}
Equivalently, the disorder-averaged Euclidean partition function takes the form
\begin{equation}
\label{Path Integral}
    \langle Z(\bj, s) \rangle_J = \int [DGD\Sigma] \, e^{-NI[G,\Sigma]} \, ,
\end{equation}
with the bilocal action functional
\begin{equation}
\label{Action}
    I=-\frac{1}{2}\log{\det(\delta(\tau_1-\tau_2)\partial_{\tau_2}-\Sigma)}+\frac{1}{2}\int_{0}^{\beta} d\tau_1\int_{0}^{\beta}d\tau_2 \left[\Sigma G-\mathcal{J}^2 \left(\frac{2^{q-1}}{q^2}G^{q}+s^2\frac{2^{\tilde{q}-1}}{\tilde{q}^2}G^{\tilde{q}} \right) \right] \, .
\end{equation}
Extremising this action 
yields the deformed Schwinger-Dyson equations \eqref{Schwinger-Dyson} as equations of motion of the large-$N$ theory. By finding solutions $(G_*,\Sigma_*)$ to these equations, one can then extract thermodynamic properties of the system by computing the on-shell action. In particular, the free energy $F$, the entropy $S$, the energy $E$, and the specific heat $C$ are given by
\begin{equation}
\label{thermodynamics}
\begin{cases}
    \beta F =N I[G_*,\Sigma_*] \,, \\
    S = (1-\beta \partial_{\beta})(-\beta F) \, , \\
    E = \partial_{\beta}(\beta F) \, , \\
    C = -\beta^2\partial^{2}_{\beta}(\beta F) \, .
\end{cases}    
\end{equation}

\noindent
That being said, solving the Schwinger-Dyson equations analytically is generally not possible. For finite values of $q$ and $\tilde{q}$, these integro-differential equations can be solved using numerical algorithms involving either the Fourier transform \cite{Maldacena:2016hyu} or Legendre polynomials \cite{Cruz:2022uic}. In what follows, with the exception of section \ref{sec: Finite q}, we do not apply such methods. Instead, we gain computational control by considering the limit in which, after taking $N$ to be large, we take both $q$ and $\tilde{q}$ to infinity.\footnote{Note that this is different from the double-scaling limit of SYK, in which both $N$ and $q$ are taken to infinity with the ratio $q^2/N$ fixed \cite{Berkooz:2024lgq}. Similar deformations to the ones explored in this paper, but in the double-scaled regime, can be found in, for instance, \cite{Berkooz:2024evs, Berkooz:2024ofm, Berkooz:2024ifu}.}

\subsection{The large $q,\tilde{q}$ limit}
Consider the limit in which $q,\tilde{q}\rightarrow \infty$ with their ratio $n \equiv q/\tilde{q} \geq 1$ fixed. In this limit, the fermion two-point function takes the form
\begin{equation}
\label{Large-q 2pt fn}
    G(\tau_1,\tau_2)=\frac{\mathrm{sgn}{(\tau_1-\tau_2)}}{2} \left(1+\frac{g(\tau_1,\tau_2)}{q}+\mathcal{O}\!\left(\!\frac{1}{q^2}\!\right) \right) \, .
\end{equation}
The action \eqref{Action} becomes 
\begin{equation}
    \label{large-q action}
    I[g] = -S_0^{\text{free}}+\frac{1}{8q^2}\int_{0}^{\beta}d\tau_1 \int_{0}^{\beta} d\tau_2 \left[\frac{1}{2}\partial_{\tau_1}g\partial_{\tau_2}g-2\mathcal{J}^2 (n^2s^2e^{g/n}+e^g) \right] \, ,
\end{equation}
 where $S_0^{\text{free}}=\log{2}/2$ is the entropy of an individual free fermion. In what follows, we work with a dimensionless time variable $t \equiv (\tau_1-\tau_2)/\beta$. With this choice, the Schwinger-Dyson equations \eqref{Schwinger-Dyson} simplify into a single ordinary differential equation,
\begin{equation}
\label{ODE}
\partial_{t}^2g(t)=2(\beta \mathcal{J})^2 \left(ns^2e^{g(t)/n}+e^{g(t)} \right) \,,
\end{equation}
supplemented by periodic boundary conditions $g(0)=g(1)=0$. Given some values for $n, s,$ and $\bj$, this equation can, in principle, be solved numerically using standard shooting methods. 

\textbf{Thermodynamic quantities.} Following \cite{Anninos:2022qgy}, the large $q,\tilde{q}$ free energy is given by 
\begin{equation}
\label{Free Energy}
    \frac{\beta F}{N}=-S_0^{\text{free}}-\frac{1}{8q^2} \int_{0}^{1} dt \left[ \frac{1}{2}(\partial_tg(t))^2+2(\beta \mathcal{J})^2 \left(n^2s^2e^{g(t)/n}+e^{g(t)}\right)\right] \, .
\end{equation}
Using \eqref{thermodynamics}, one can then derive the thermodynamic entropy,
\begin{equation}
\label{Old Entropy}
    \frac{S}{N}=S_0^{\text{free}}+\frac{1}{8q^2} \int_{0}^{1}dt \left[\frac{1}{2}(\partial_tg(t))^2-2(\beta \mathcal{J})^2 \left(n^2s^2e^{g(t)/n}+e^{g(t)}\right) \right] \,,
\end{equation}
and the energy,
\begin{equation}
\label{Energy}
    \frac{E}{N} = - \frac{\beta \mathcal{J}^2}{2q^2} \int_{0}^{1} dt \left(n^2 s^2 e^{g(t)/n}+e^{g(t)}\right) \, .
\end{equation}
In this work, we will be mostly interested in characterising the behaviour of the entropy at low temperatures. In principle, we could proceed in the following manner: solve \eqref{ODE} numerically for some values of $n, s$, and $\beta \mathcal{J}$, plug the solution $g(t)$ into \eqref{Old Entropy}, and numerically integrate to find the entropy, then repeat the procedure for other values of the parameters. However, this method is particularly inefficient (and inaccurate) at very low temperatures. 

Instead, note that the equations of motion \eqref{ODE} render the integrand in \eqref{Old Entropy} time-independent. The $t \rightarrow 1-t$ symmetry of the problem then simplifies the entropy to
\begin{equation}
\label{Entropy}
    \frac{S}{N}=S_0^{\text{free}}-\frac{(\beta \mathcal{J})^2}{4q^2} \left(n^2s^2e^{g_m/n}+e^{g_m}\right) \,,
\end{equation}
where $g_m \equiv g(t=1/2)$. One way to obtain $g_m$ is to numerically solve \eqref{ODE} for the whole two-point function. Alternatively,  we simply observe that \eqref{ODE} can be solved, at least formally, by
\begin{equation}
    \label{integral solution}
    -2\beta \mathcal{J} \left(t-\frac{1}{2} \right) = \int_{g_m}^{g(t)} \frac{dx}{\sqrt{W(x)-W(g_m)}} \,,
\end{equation}
 over $0 \leq t \leq 1/2$, where $W(x) \equiv n^2s^2e^{x/n}+e^x$. Therefore,
\begin{equation}
    \label{gstar integral}
    \beta \mathcal{J} = \int_{g_m}^{0} \frac{dx}{\sqrt{W(x)-W(g_m)}} \, \, .
\end{equation}
Once integrated, \eqref{gstar integral} can be inverted to find $g_m$ as a function of $\beta \mathcal{J}$, $n$, and $s$, and therefore the entropy. Note that this method of computing the entropy does not require knowledge of $g(t)$ in the entire range of $t$, but rather just its value $g_m$ at $t=1/2$. In what follows, we will make strong use of this approach (both analytically and numerically), which allows us to compute the thermodynamic entropy for a wide range of parameters, particularly in the low-temperature regime.

\textbf{Analytically solvable examples.} 
Two cases have been previously solved analytically in the literature.

The first is the undeformed SYK model. For $n=1$, \eqref{ODE} is simply the equation of motion for a single SYK model $H_q$ in the large-$q$ limit with the coupling rescaled by $\J \to \J \sqrt{1+s^2}$. In this case, the analytical solution at finite temperature is given by
\begin{equation}
\label{n=1 solution}
    e^{g(t)} = \frac{\cos^2(\nu)}{\cos^2 \left(2\nu \left(t-\frac{1}{2}\right)\right)} \, \, \, \, , \, \, \, \bj=\frac{2\nu}{\sqrt{1+s^2}\cos(\nu)} \, ,
\end{equation}
from which one obtains $e^{g_m} = \cos^2(\nu)$. At low temperatures, this reduces to
\begin{equation}
\label{n=1 gstar}
    e^{g_m} = \frac{\pi^2}{1+s^2}\frac{1}{(\bj)^2}-\frac{4\pi^2}{(1+s^2)^{3/2}}\frac{1}{(\bj)^3}+\mathcal{O}\!\left(\!\frac{1}{(\beta \mathcal{J})^4}\!\right) \,.
\end{equation}
By inserting this low-temperature expansion into \eqref{Entropy}, we obtain
\begin{equation}
    \label{n=1 entropy}
    \frac{S}{N} = S_0^{\text{free}}-\frac{\pi^2}{4{q}^2}+\frac{1}{\sqrt{1+s^2}}\frac{\pi^2}{q^2}\frac{1}{\beta \mathcal{J}}+\mathcal{O}\!\left(\! \frac{1}{(\beta \mathcal{J})^2} \!\right) \, ,
\end{equation}
which is, after the shift in the coupling, the celebrated linear-in-temperature entropy of the SYK model at strong coupling.

Interestingly, there is another value of $n$ for which \eqref{ODE} can be solved exactly. This is the case of $n=2$, which has been extensively analysed in \cite{Jiang:2019pam,Anninos:2022qgy}. The exact analytical two-point function for any value of $\bj$ and $s$ is given by
\begin{equation}
    \label{n=2 solution}
    e^{g(t)/2}=\frac{2\mu^2}{ \sqrt{(\beta \mathcal{J})^2\mu^2+(\beta \mathcal{J})^4 s^4} \cos \left(2\mu \left(t-\frac{1}{2}\right)\right) + (\beta \mathcal{J})^2s^2} \, \, \, , \, \, \, \cos(\mu)=\frac{2\mu^2-(\bj)^2s^2}{\sqrt{(\bj)^2 \mu^2+(\bj)^4s^4}} \, ,
\end{equation}
from which we can infer the value of $g_m$,
\begin{equation}
\label{n=2 gstar}
    e^{g_m/2} = \frac{2\mu^2}{\sqrt{(\beta \mathcal{J})^2\mu^2+(\beta \mathcal{J})^4 s^4}  + (\beta \mathcal{J})^2s^2} \,.
\end{equation}
We are interested in the behaviour of the system at low temperatures $\bj \gg 1$. In this regime, two distinct parametric regions exist, which we refer to as the intermediate IR ($1 \ll \bj \ll 1/s^2$) and the deep IR ($\bj \gg 1/s^2$). Note that in order for the intermediate IR to appear, we need $s^2\ll1$. In both cases, $g_m$ admits a systematic expansion, 
\begin{equation}
\label{n=2 gstar cases}
    e^{g_m/2} = \begin{cases}
        \frac{\pi}{\bj}-\frac{2\left(\pi^2+(\pi-2)(\bj s)^2\right)}{\pi(\bj)^2} + \mathcal{O}\!\left(\!\frac{1}{(\bj)^3}\!\right) \, , \, & \text{Intermediate IR} \,, \\
        \frac{\pi^2}{s^2(\bj)^2}-\frac{2\pi^2 \sqrt{1+4s^2}}{s^4(\bj)^3} + + \mathcal{O}\!\left(\!\frac{1}{(\bj)^4}\!\right) \, , \, & \text{Deep IR} \,,
    \end{cases}
\end{equation}
which yields the low-temperature entropy,

\begin{equation}
    \label{n=2 entropy}
    \begin{aligned}
        \text{Intermediate IR}: \ \ \frac{S}{N} &= S_0^{\text{free}}-\frac{\pi^2}{4q^2}+\frac{\pi^2}{q^2}\frac{1}{\beta \mathcal{J}}-\frac{2\beta \mathcal{J}s^2}{q^2}+ \mathcal{O}\!\left(\!\frac{1}{(\bj)^2}\!\right) \, , \\
        \text{Deep IR}: \ \ \frac{S}{N} &= S_0^{\text{free}}-\frac{\pi^2}{4\tilde{q}^2}+\frac{\sqrt{1+4s^2}}{2s^2}\frac{\pi^2}{\tilde{q}^2}\frac{1}{\beta \mathcal{J}}+\mathcal{O}\!\left(\!\frac{1}{(\bj)^2}\!\right) \, .
    \end{aligned}
\end{equation}
The entropy in the intermediate IR can be identified with the entropy \eqref{n=1 entropy} of the single $n=1$ SYK model in the large-$q$ limit, with $s=0$. In the deep IR, the entropy also has a linear-in-temperature behaviour. However, the coefficient is {\textit{not}} the one expected for a single SYK model. Instead, in \cite{Anninos:2022qgy}, this deformation has been interpreted in terms of a renormalisation group (RG) flow at finite temperature, in which the term $sH_{\tilde{q}}$ in the Hamiltonian acts as a relevant deformation of the single SYK model with $q$-fermion interactions.

Away from these analytical values of $n$, the idea of the deformation inducing an RG flow has been explored numerically both at finite and infinite $q,\tilde{q}$ in \cite{Anninos:2022qgy}. The analysis has been performed mostly for $n\geq 2$. In this work, we further extend the results to lower values of $n$. Interestingly, we find that for $1<n<3/2$, new \textit{anomalous} behaviour emerges for the entropy as a function of temperature in the deep IR. Moreover, for exactly $n=3/2$, there is a logarithmic enhancement of the linear-in-temperature behaviour. Following the method described above, we find analytical results for the entropy in the deep IR for a few particular values of $n$. Furthermore, we analytically determine the regions of validity of the intermediate and deep IR, for all values of $n$, clarifying previous confusions regarding the zero-temperature behaviour of the model for $1<n<2$. The model can also be solved perturbatively around both $n=1$ and $n=2$, which we show in appendix \ref{sec: exact appendix}.

\section{Deep infrared thermodynamics} \label{sec: DeepIR}

We start by characterising the deep infrared regime of the model, which is achieved at low temperatures, much lower than any scale set by $s$. The intuition, gained from the analytically solvable $n=2$ model, is that a new parametric regime of linear-in-temperature behaviour emerges in this limit. In \cite{Anninos:2022qgy}, there was further numerical and analytical evidence that this linear regime would hold for other values of $n\geq 2$. For $n<2$, the numerical evidence provided in \cite{Anninos:2022qgy} showed a linear-in-temperature behaviour at the cost of having a zero-temperature entropy that depended on $s$. We will show that this picture needs to be reconsidered. 

The first result of this section is that in the deep infrared, the deformed SYK model has, at large $N$, the usual zero-temperature entropy corresponding to a pure SYK model. The leading correction away from zero temperature depends on the value of $n$. For $n>3/2$, the model has linear-in-temperature entropy in the deep IR. For $n<3/2$, the leading contribution has an anomalous scaling that goes as $(\bj)^{2-2n}$ which, in this regime, dominates over any linear-in-temperature term. At exactly $n=3/2$, the power-law behaviour is corrected to $\tfrac{\log \bj}{\bj}$.

\begin{figure}[!htbp]
    \centering
    \includegraphics[width=0.6\textwidth]{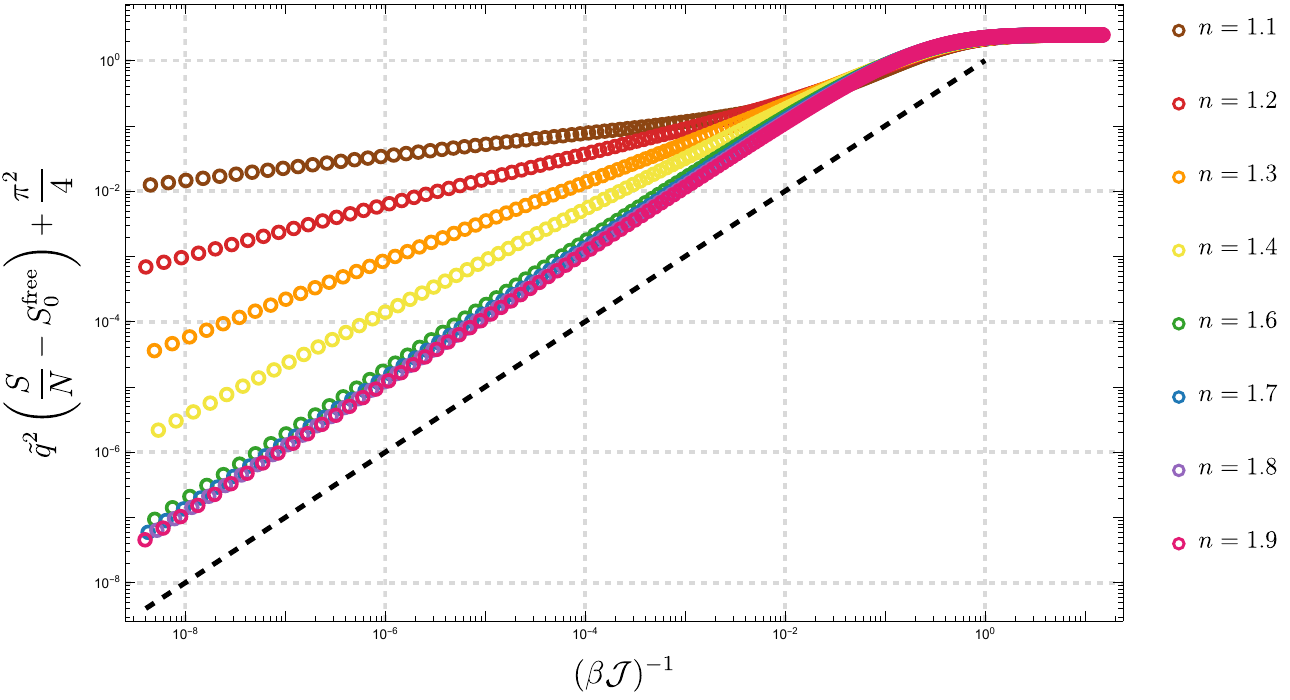}
    \caption{Entropy as a function of temperature for different values of $n$. The circles are numerical values of $\tilde{q}^2\!\left(\frac{S}{N}-S_0^{\text{free}} \right)+\frac{\pi^2}{4}$ as a function of $(\bj)^{-1}$ for $s=1$. For reference, the black dashed line corresponds to exactly linear in temperature entropy.}   \label{fig: transition plot}
\end{figure}

We start by providing numerical evidence for this scaling. We fix $s=1$ and first solve \eqref{gstar integral} to obtain $g_m$ at low temperatures. 
Then we plug $g_m$ into \eqref{Entropy} to find the entropy. We plot these numerical results in figure \ref{fig: transition plot}, where the power-law behaviour becomes evident. It is also clear that, for $n<3/2$, the scaling of the entropy stops being linear. It is important to note that for small values of $s$ and values of $n$ close to 1, the deep IR regime emerges at lower temperatures, making it difficult to access with the usual numerical techniques. Our approach allows us to access these very low temperatures ($\bj > 10^{15}$). We derive the exact parametric regime of validity in the next section.

\subsection{Zero-temperature entropy} \label{sec: zero-temp entropy} 
\begin{figure}[!htbp]
    \centering
    \includegraphics[width=0.6\textwidth]{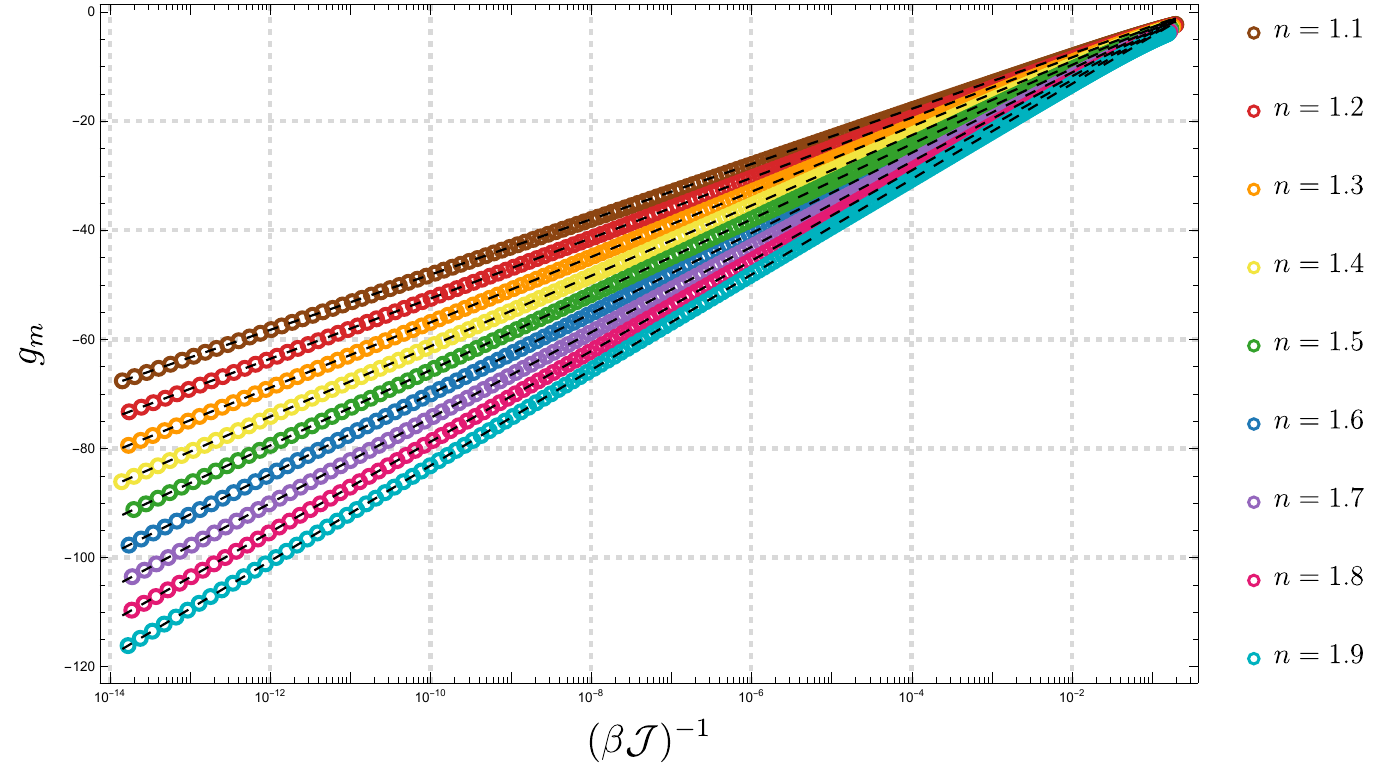}
    \caption{The value of the two-point function $g(t=1/2)$ as a function of temperature. The circles are numerical values of $g_m$ as a function of $(\bj)^{-1}$ for $s=1$ and various values of $n$. The dashed lines correspond to the analytical prediction $g_m = 2n \log\! \left(\!\frac{\pi}{\bj s}\!\right)$.}    \label{fig: zero-temperature graph}
\end{figure}

Here, we provide an analytical argument for the zero-temperature entropy of the model. The strategy is to study the equation of motion \eqref{ODE} perturbatively at low temperatures, identify the low-temperature behaviour of $g_m$, and determine the entropy through \eqref{Entropy}. For convenience, we recall \eqref{ODE} here:
\begin{equation}
\label{eq: ODE again}
\partial_{t}^2g(t)=2(\beta \mathcal{J})^2 \left(ns^2e^{g(t)/n}+e^{g(t)} \right) \, .
\end{equation}
Given the $(\bj)^2$ on the right side of this equation, the only non-trivial leading term at low temperatures takes the form
\begin{equation}
\label{leading g}
    e^{g(t)/n} = \frac{g_0(t)}{(\bj)^2} + \cdots,
\end{equation}
where $g_0(t)$ is a smooth function that does not depend on $\bj$, and the dots correspond to terms that are subleading as $\bj \to \infty$. As expected, by plugging this into \eqref{eq: ODE again}, all the $\bj$-dependence drops, and we are left with an ordinary differential equation for $g_0(t)$,
\begin{equation}
 \partial_t^2 g_0(t) = \frac{(\partial_t g_0(t){})^2}{g_0(t)} + 2 s^2 g_0(t){}^2 \,.
\label{eq: g0 ODE}
\end{equation}
Importantly, as we are taking the low-temperature limit first, we cannot trust this approximation when $t$ or $(1-t)$ are of the order $(\bj)^{-1}$. Accordingly, we cannot trivially inherit the periodic boundary conditions at $t=0,1$. Instead, the $t\to 1-t$ symmetry, which imposes $\partial_t g_0(t=1/2) = 0$, partially constrains the solution. In particular, we have
\begin{equation}
    \label{g0 with unknown, sec3}
    g_0(t)=\frac{c_0^2}{4s^2 \cos^2\left(\frac{c_0}{4}(2t-1) \right)} \,,
\end{equation}
for some constant $c_0$. This constant can be fully determined using an appropriate limit, which is described in appendix \ref{sec: appendix A}. The final result is
\begin{equation}
    \label{y0 and y0 tilde}
    |c_0|=2\pi \, \, \longrightarrow \, \, g_0(t=1/2) = \frac{\pi^2}{s^2} \, \quad , \quad \text{for all $n > 1$} .
\end{equation}
This result is numerically verified in figure \ref{fig: zero-temperature graph}. Plugging this into \eqref{leading g} and subsequently into \eqref{Entropy}, we find that the zero-temperature entropy in the deformed SYK model is, for all $n > 1$, given by
\begin{equation}
\label{zero-temp entropy}
  \left.  \frac{S}{N} \right|_{\bj \to \infty} = S_0^{\text{free}}-\frac{\pi^2}{4\tilde{q}^2} \,.
\end{equation}
In particular, we observe that there is no dependence on $s$.  Note that this contrasts with figure 1 in \cite{Anninos:2022qgy}. The reason is that the temperatures used in that numerical study are not really in the deep IR regime for values of $n \gtrsim 1$. We will come back to this in section \ref{sec: intermedIR}.

\subsection{Low-temperature entropy}
\label{low-temperature entropy}

We now proceed with the analysis of the low-temperature entropy away from the zero-temperature point. Following the same procedure, it can be shown that $g(t)$ admits a systematic, self-consistent expansion at low temperatures, given by
\begin{equation}
    \label{eq: generally g}
    e^{g(t)/n} = \sum_{k=0}\frac{g_k(t)}{(\bj)^{k+2}}+\sum_{k=1}\frac{h_k(t)}{(\bj)^{2+2(n-1)k}} + \sum_{k=1}\frac{\ell_{k}(t)}{(\bj)^{2n+k}}\, \, ,
    \end{equation}
where $g_k(t)$, $h_k(t)$, and $\ell_{k}(t)$ are smooth functions satisfying $\partial_t g_k(t=1/2) = \partial_t h_k(t=1/2)=\partial_{t}\ell_{k}(t=1/2)=0$. As we will see below, the contribution of the $g_1(t)$ term to the entropy is linear in temperature. As the $\ell_k(t)$ terms (and potentially others not present in the above expansion) are always subleading compared to the contribution of the $g_1(t)$ term, they will not be analysed here. 

Note that this expansion holds for generic values of $n$. However, for particular values, such as those of the form $1+\frac{1}{2k}$ with $k \in \mathbb{N}^{*}$, the entropy may have additional contributions which are logarithmic in temperature. We will specifically study the case of $n=3/2$ in section \ref{sec: more analytically solvable cases}. For $n \in \mathbb{N}^{*}$, instead, only the sum with $g_k(t)$ exists.

Given $n>1$, the leading term for every $n$ will always be the same $g_0(t)$ from \eqref{g0 with unknown, sec3} with $c_0=2\pi$. It is clear that the temperature-dependence of the subleading term differs depending on the value of $n$. More specifically, it is given by $h_1(t)$ for $n<3/2$, and $g_1(t)$ for $n>3/2$. Plugging \eqref{eq: generally g} into \eqref{eq: ODE again}, we find
\begin{equation}
    \label{Perturbative ODEs}
    \begin{cases}
        \partial_{t}^2g_1(t)=-4\pi \cot(\pi t)\partial_{t}g_1(t)+4\pi^2g_1(t) \, , \\
        \partial_{t}^2h_1(t)=-4\pi \cot(\pi t)\partial_{t}h_1(t)+4\pi^2h_1(t)+\frac{2}{n}\left(\frac{\pi}{s \sin(\pi t)} \right)^{2n+2} \, ,
    \end{cases}
\end{equation}

after plugging in the solution for $g_0(t)$. Once again, we have assumed that $n$ is generic. These equations admit analytical solutions,
\begin{equation}
    \label{Solution 1 to perturbative ODEs}
    \begin{cases}
        g_1(t) = \frac{c_1(2+\pi(1-2t)\cot(\pi t))}{\sin^2(\pi t)} \, , \\
 h_1(t) = \frac{c_2 (\pi  (2 t-1) \cot (\pi  t)-2)}{\pi \sin^2(\pi t) } -\frac{\pi ^{2 n}}{n^2s^{2n+2}\sin^{2n}(\pi t)}
        +\frac{2 (n-1) \pi ^{2 n}  \cos ^2(\pi  t) \, _2F_1\left(\frac{1}{2},n-\frac{1}{2};\frac{3}{2};\cos ^2(\pi  t)\right)}{n^2s^{2n+2}\sin^3(\pi t)} \,,
        \end{cases}
\end{equation}

where we used the boundary conditions $\partial_t g_1(t=1/2) = \partial_t h_1(t=1/2)=0$, and $c_1,c_2$ are constants of integration (which may depend on $n$ and $s$) to be specified by an additional boundary condition. We will see shortly that $h_1(t)$ can be completely determined following the argument in appendix \ref{appendix A zero-temp entropy}. For now, let us observe the leading low-temperature behaviour of the entropy.

For $n>3/2$, by plugging \eqref{eq: generally g} into \eqref{Entropy}, we find
\begin{equation}
\label{entropy n>3/2}
    n > \frac{3}{2}: \qquad \frac{S}{N} = S_0^{\text{free}}-\frac{\pi^2}{4\tilde{q}^2}+\frac{S_1 (n,s)}{\tilde{q}^2}\frac{1}{\beta \mathcal{J}}+\cdots , \\
\end{equation}
where we have assumed that $g_1(t=1/2) \neq 0$. This assumption is supported by numerical evidence, see below. Therefore, when $n>3/2$, the leading correction away from zero temperature is linear in temperature. This is the standard Schwarzian-dominated entropy which was found both numerically and via conformal perturbation theory in \cite{Anninos:2022qgy}. Once again, the next correction depends on $n$. For $n \geq 2$, the next correction scales as $(\bj)^{-2}$ while for $3/2 < n < 2 $, it goes as $(\bj)^{2-2n}$.

Interestingly, the situation changes qualitatively when $n<3/2$. In that case, we find that the entropy in the deep infrared is given by
\begin{equation}
     1<n<\frac{3}{2}: \qquad \frac{S}{N} = S_0^{\text{free}}-\frac{\pi^2}{4\tilde{q}^2}+\frac{\tilde{S}_1 (n,s)}{\tilde{q}^2}\frac{1}{(\beta \mathcal{J})^{2n-2}}+ \cdots \, , \label{eq: entropy n<3/2}
\end{equation}
where $\tilde{S}_1(n,s)$ is some function assumed to be non-zero, see below for analytical and numerical confirmation. In this case, the leading term in \eqref{eq: entropy n<3/2} dominates over the standard linear-in-temperature entropy of the SYK model. This is consistent with the numerical results found in figure \ref{fig: transition plot}. The dots indicate subleading corrections that differ depending on the value of $n$. The next term is linear in temperature if $5/4<n < 3/2$, while it goes as $(\bj)^{4-4n}$ for $1<n<5/4$.\footnote{As $n$ gets closer to 1, more and more of the $h_k(t)$ terms in \eqref{eq: generally g} dominate over the $g_1(t)$ term at the lowest temperatures. The effect on the entropy is that there may be more contributions that dominate over the linear-in-temperature term.}

From the renormalisation group flow perspective, starting from the near-conformal theory in the deep infrared, we can interpret the leading deformation in the anomalous cases $1<n<3/2$ as an operator of dimension $\Delta = n$ that dominates the Schwarzian. This operator is still there when $n>3/2$, except its contribution to the entropy is dominated by that of the Schwarzian. The same scaling was observed in a theory with two coupled SYK models \cite{Milekhin:2021cou}. It would be interesting to explore the connection between the two models.

Note that the appearance of this anomalous scaling at low temperatures will also be the generic case at finite $q$, $\tilde q$. It is only the $q=4$ model that does not have such operators. The $q=6$ model has a $\tilde q=4$ deformation with $\Delta = 3/2$, and for any $q>6$, there exists at least one operator with $1<\Delta<3/2$. We provide some numerical results of the model at finite $q$ in section \ref{sec: Finite q}. 

In short, we have identified a family of deformed SYK models where the Schwarzian does not dominate in the infrared \cite{Maldacena:2016upp}. We comment on how this result is consistent with expectations from conformal perturbation theory in appendix \ref{sec: appendix on conformal perturbation theory}, following the procedure described in \cite{Anninos:2022qgy}.

\textbf{Dependence on $n$ and $s$.} Having computed the leading low-temperature scaling of the deformed SYK models, we now discuss the dependence on $n$ and $s$ in the coefficients $S_1$ and $\tilde{S}_1$ in \eqref{entropy n>3/2} and \eqref{eq: entropy n<3/2} respectively.

\begin{figure}[!htbp]
    \centering

    \subfigure[$\tilde{S}_1(n,s)$]{
        \includegraphics[width=0.48\textwidth]{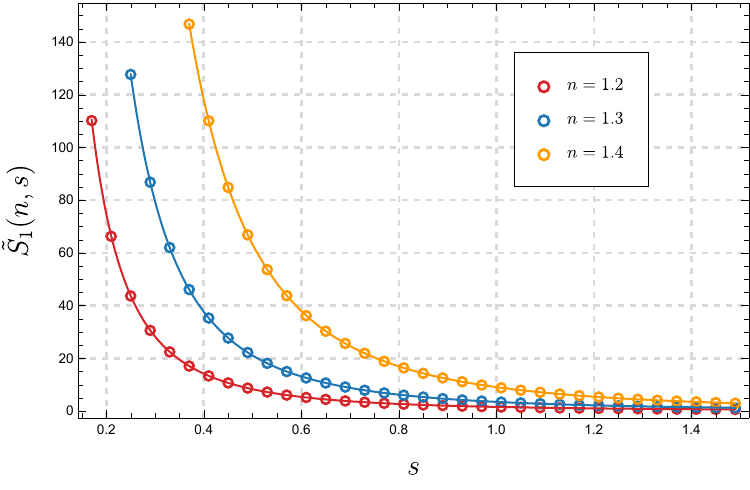}
        \label{fig: anomalous coef}
    }
    \hfill
    \subfigure[$S_1(n,s)$]{
        \includegraphics[width=0.48\textwidth]{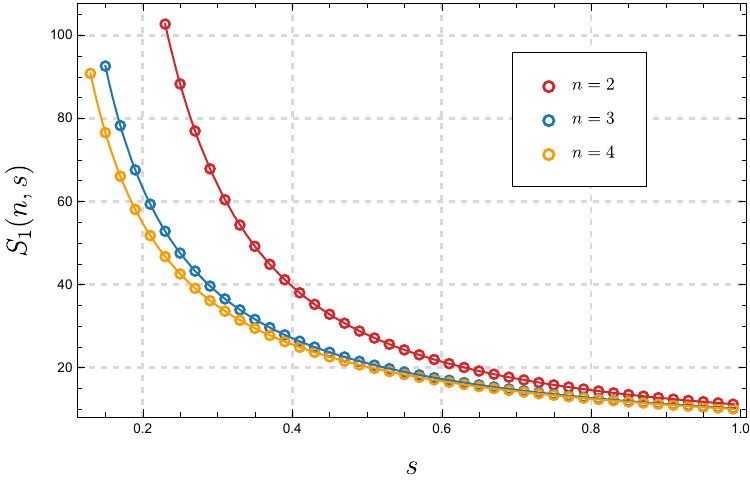}
        \label{fig: analytical linear coefs}
    }

    \caption{The functions (a) $\tilde{S}_1(n,s)$ and (b) $S_1(n,s)$ as a function of $s$, for different values of $n$. The circles correspond to numerical evaluations while the solids lines are analytical results from \eqref{eq: tilde S1} in (a) and \eqref{linear coef n=2}, \eqref{Linear coef n=3}, and \eqref{Linear coef n=4} in (b).}
    \label{fig:analytical results}
\end{figure}

We start by computing $\tilde S_1$. Numerically, this can be computed following the procedure described in section \ref{sec: deformedSYK}. Namely, we first find $g_m$ using \eqref{gstar integral}, plug it in the entropy formula \eqref{Entropy}, subtract the zero-temperature entropy \eqref{zero-temp entropy}, and obtain $\tilde S_1$. Remarkably, $\tilde S_1$ can also be computed analytically for any value of $n$. Following the method described in appendix \ref{appendix A zero-temp entropy}, we first determine $c_2$ in \eqref{Solution 1 to perturbative ODEs}. This directly establishes a closed analytical form for $\tilde S_1$, 
\begin{equation}
 c_2 = \frac{(2 n-1) \pi ^{2 n+\frac{1}{2}}  \Gamma \left(\frac{1}{2}-n\right)}{2 n^2 \Gamma (1-n)}\frac{1}{s^{2n+2}} \quad \longrightarrow \quad    \tilde S_1 (n,s) = (2n-1) \frac{ \pi ^{2 n-\frac{1}{2}} \Gamma\!\left(\frac{1}{2}-n\right)}{4n^2 \Gamma (1-n)}\frac{1}{s^{2n}} \, \, ,
    \label{eq: tilde S1}
\end{equation}
which agrees with the numerical results to excellent precision, as shown in figure \ref{fig: anomalous coef}. Note that this result remains valid even when this anomalous term is subleading compared to the linear-in-temperature term, which is the case when $n > 3/2$. Furthermore, \eqref{eq: tilde S1} matches with the leading correction to the thermal entropy one would obtain from conformal perturbation theory, provided the conformal theory is deformed by an operator of dimension $\Delta = n$, and whose coupling is given by
\begin{equation}
    \label{deep IR coupling}
    g_{\Delta=n}^2=\frac{\mathcal{J}^2}{2n^2s^{2n}\tilde{q}^2}\ \, .
\end{equation}
A derivation of the conformal perturbation theory result is given in appendix \ref{sec: appendix on conformal perturbation theory}, following \cite{Tikhanovskaya:2020elb, Cruz:2022uic}.

Next, we turn to computing $S_1(n,s)$, the coefficient of the linear-in-temperature portion of the entropy. For the moment, we restrict the analysis to the cases in which this contribution dominates at very low temperatures, that is, $n>3/2$. Unlike $\tilde S_1$, we do not have an analytical formula for $S_1$, except for particular values of $n$. One such case is $n=2$, described in section \ref{sec: deformedSYK}, where
\begin{equation}
\label{linear coef n=2}
    S_1 (n=2, s)= \frac{\pi^2 \sqrt{1+4s^2}}{2s^2} \,.
\end{equation}
Two more cases where we can find an analytical expression for $S_1(n,s)$ are $n=3$ and $n=4$, see appendix \ref{sec: appendix other n}. We use these two analytical values to check that our numerical scheme gives appropriate results, and find excellent agreement between the analytical results and the numerical evaluations in figure \ref{fig: analytical linear coefs}.

\begin{figure}[!htbp]
    \centering

    \subfigure[]{
        \includegraphics[width=0.48\textwidth]{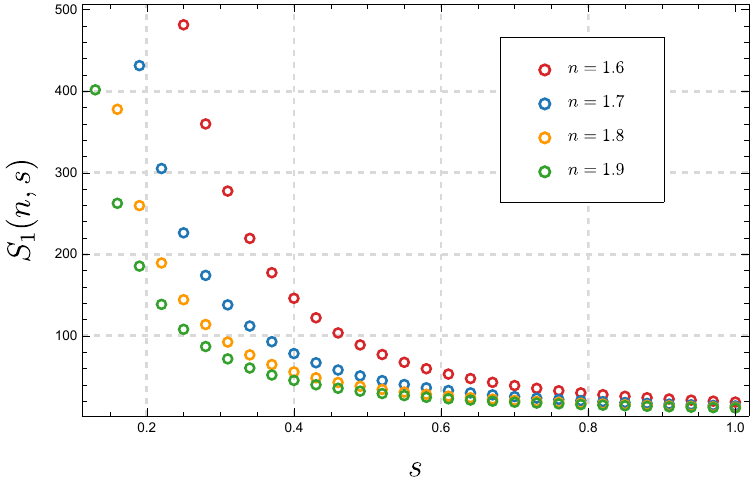}
        \label{fig:numerical_linear_coef_vs_s}
    }
    \hfill
    \subfigure[]{
        \includegraphics[width=0.48\textwidth]{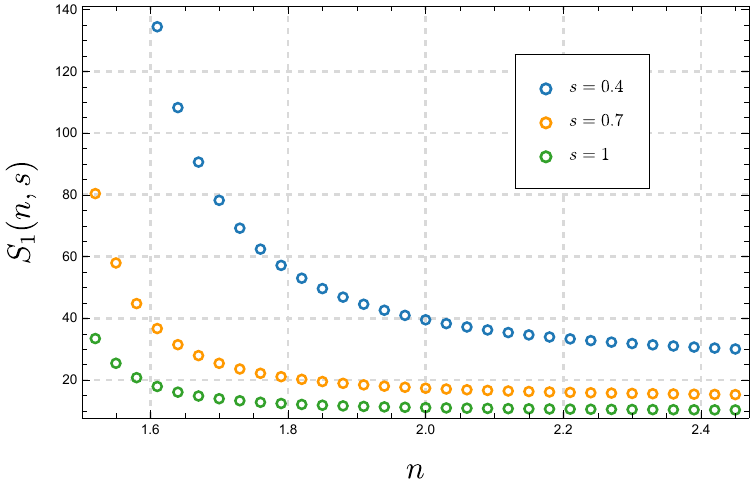}
        \label{fig:numerical_linear_coef_vs_n}
    }

    \caption{The coefficient $S_1$ as a function of both $n$ and $s$. In figure (a), we show the results for different values of $n$, as a function of $s$, while in figure (b), we show the results for different values of $s$, as a function of $n$. Circles correspond to numerical evaluations.}
    \label{fig:numerical_linear_coef}
\end{figure}

For general values of $n$ and $s$, we can find the coefficient $S_1$ numerically, which we show in  figure \ref{fig:numerical_linear_coef}. This numerical data is consistent with two other limits that can be solved analytically, namely,
\begin{equation}
    S_1(n \rightarrow \infty,s) = S_1(n,s \r \infty) = \frac{\pi^2}{s} \,.
\end{equation}
Furthermore, from these numerical explorations, it seems that for $n>3/2$, $S_1(n,s)$ at small $s$ behaves as
\begin{equation}
    \label{small s linear coef}
    S_1(n,s \ll 1) = \frac{a(n)}{s^{\frac{n}{n-1}}}+\cdots \, .
\end{equation}
The dots correspond to terms that are subleading at small $s$. In order to see this functional dependence, we needed to take $s \approx10^{-6}$. Further details, including a numerical plot of the function $a(n)$, can be found in appendix \ref{sec: S1 at small s appendix}.


\subsection{More analytically solvable cases}

\label{sec: more analytically solvable cases}

While \eqref{ODE} cannot be solved for generic values of $n$, we are still capable of finding exact coefficients of the low-temperature entropy for a few very particular values of $n$. The idea is to invert \eqref{gstar integral} to obtain $g_m$ as a function of $\bj$, and then use \eqref{Entropy} to find the low-temperature entropy.

Assume that $n$ takes a rational value $n=\tfrac{u}{v}$ with $u>v \in \mathbb{Z^+}$. Then, after the change of variable $y=e^{x/u}$, \eqref{gstar integral} becomes
\begin{equation}
    \label{gstar integral again}
    \beta \mathcal{J} = \int_{e^{g_m/u}}^{1} \frac{u \, dy}{y\sqrt{y^u + n^2s^2 y^v-\left(e^{g_m}+n^2s^2 e^{g_m/n}\right)}} \, \, .
\end{equation}
Integrals of this form can be rewritten in terms of elliptic integrals as long as the degree $u$ of the polynomial under the square root is at most 4 \cite{Byrd:1971bey}. In our case, given that $u>v$, this translates into the following very few cases in which we might hope to obtain closed expressions for $\bj$ in terms of $g_m$,
\begin{equation}
    \label{solvable cases}
 \text{Analytically tractable values of $n$:} \quad    \bigg\{1, \,2, \,3, \,4, \,\frac{3}{2}, \,\frac{4}{3}\bigg\} \, .
\end{equation}
Even if we formally obtain $\bj$ as a function of $g_m$ for these particular values, inverting the expression may be complicated. The two simplest cases are $n=1$ and $n=2$, which were discussed in section \ref{sec: deformedSYK}. In these cases, the integral can be written in terms of elementary functions. For the remaining cases, we may, after performing the integral, apply a perturbative expansion in $e^{g_m/u} \ll 1$ in order to invert the expression order by order and obtain the low-temperature expansion of $g_m$. One can then find the low-temperature coefficients of the entropy using \eqref{Entropy}. The resultant expressions for $n=3$ and $n=4$ are complicated, but the interested reader can find them in appendix \ref{sec: appendix other n}. 

In the rest of this subsection, we focus on the two remaining cases. When $n=3/2$, there is a logarithmic enhancement of the linear-in-temperature entropy. While for $n=4/3$, we find that the subleading linear-in-temperature behaviour can, in fact, completely vanish for some particular value of the coupling $s$.

\subsubsection{The case of $n=3/2$}

For this particular value of $n$, the two leading contributions in the expansion \eqref{eq: generally g} of the two-point function at low temperatures naively overlap. However, what actually happens is that one of them gets enhanced by a logarithmic correction. After writing \eqref{gstar integral again} at $n=3/2$ in terms of elliptic integrals, inverting the expression perturbatively yields

\begin{equation}
    \begin{aligned}
        e^{\tfrac{2g_m}{3}} = &\frac{\pi ^2}{s^2 (\bj)^2}-\frac{8 \pi ^2}{9s^5}\frac{\log(\bj)}{(\bj)^3} \\
    &-\frac{4 \pi ^2 \left(3 s \sqrt{4+9 s^2}+2 \log \left(\frac{9s^3}{2\pi}\left(\sqrt{4+9 s^2}-3 s\right)^2\right)+\pi -4\right)}{9s^5 (\bj)^3} + \mathcal{O}\!\left(\frac{\log(\bj)^2}{(\bj)^4}\right) \, .
    \end{aligned}
\end{equation}

\begin{figure}[h!]
    \centering

    \subfigure[$S_{\frac{\log(\bj)}{\bj}}$]{
        \includegraphics[width=0.48\textwidth]{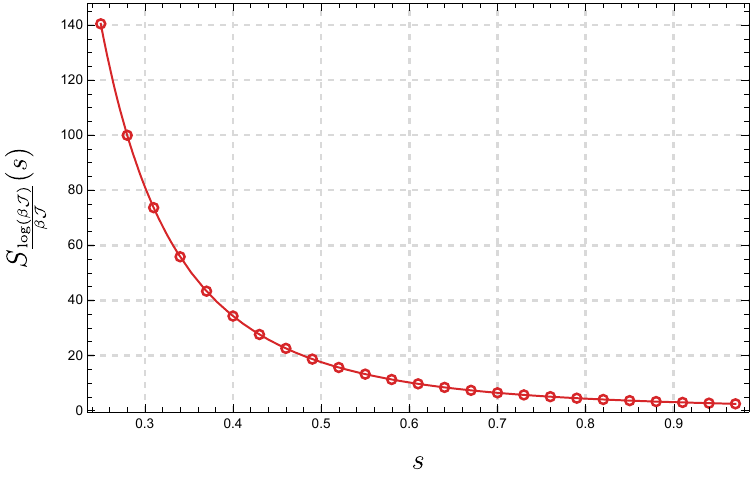}
        \label{fig:n=1.5_TlogT}
    }
    \hfill
    \subfigure[$S_{\frac{1}{\bj}}$]{
        \includegraphics[width=0.48\textwidth]{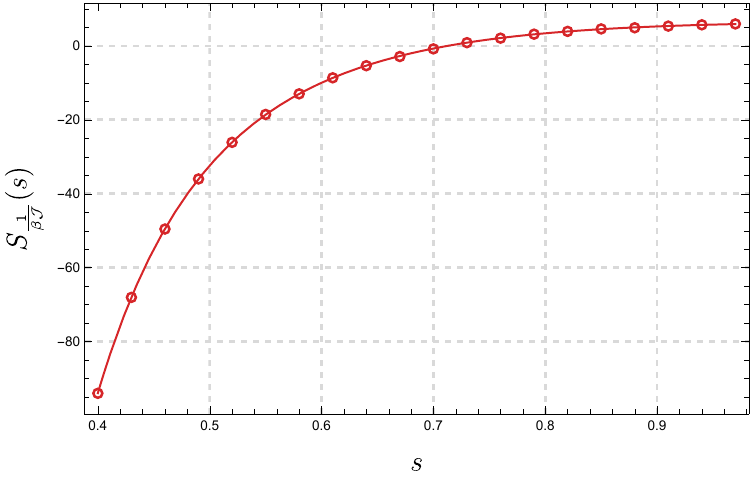}
        \label{fig:n=1.5_linear}
    }

    \caption{The coefficients of the low-temperature entropy for $n=3/2$ as a function of $s$. In figure (a), we show the coefficient of the $\frac{\log(\bj)}{\bj}$ term, while in figure (b), we show the coefficient of the linear-in-temperature term. Circles correspond to numerical evaluations while the solid lines are the analytical results in \eqref{n=1.5 entropy coefs}.}
    \label{fig:3_over_2_coefficients}
\end{figure}

Therefore, the low-temperature entropy for $n=3/2$ is given by 
\begin{equation}
    \label{n=1.5 entropy}
        n=\frac{3}{2}: \qquad \frac{S}{N} = S_0^{\text{free}}-\frac{\pi ^2}{4 \tilde{q}^2}+\frac{S_{\frac{\log(\bj)}{\bj}}}{\tilde{q}^2}\frac{\log(\beta \mathcal{J})}{\beta \mathcal{J}} + \frac{S_{\frac{1}{\bj}}}{\tilde{q}^2}\frac{1}{\bj}  + \mathcal{O}\!\left(\frac{\log(\bj)^2}{(\bj)^2}\right) \,,
\end{equation}
with
\begin{equation}
    \label{n=1.5 entropy coefs}
       {S_{\frac{\log(\bj)}{\bj}}} =  \frac{2 \pi ^2 }{9 s^3} \,, \qquad S_{\frac{1}{\bj}} =
         \frac{\pi ^2 \left(2 \log \left(\frac{9 s^3}{2 \pi } \left(\sqrt{4+9 s^2}-3 s \right)^2\right)+3 s \sqrt{4+9 s^2}-4\right)}{9 s^3} \,.
\end{equation}
These expressions can be checked numerically, as shown in figure \ref{fig:3_over_2_coefficients}, where we find excellent agreement with the numerical results. In figure \ref{fig:3_over_2_corrections} we show the low-temperature predictions against the full entropy at low temperatures for $s=0.01$, computed numerically.

\subsubsection{The case of $n=4/3$} 

This value of $n$ serves as a particular example of the anomalous deformations for which the integral \eqref{gstar integral again} is easy to invert when evaluated perturbatively. Following this method, we find the expansion of $g_m$ at low temperatures,
\begin{equation}
    \label{4/3 gstar}
        e^{\tfrac{3g_m}{4}}=  \frac{\pi ^2}{(\bj s)^2}-\frac{3 \left(3 \pi ^{8/3} \Gamma \left(-\frac{1}{3}\right)+5 \pi ^{13/6} \Gamma \left(-\frac{5}{6}\right)\right)}{16 \Gamma \left(-\frac{1}{3}\right) s^{14/3}(\bj)^{8/3}}
        -\frac{\pi ^2 \left(8 s^2-9\right) \sqrt{9+16 s^2}}{8 (\bj) ^3 s^6 (\bj)^3}+\mathcal{O}\!\left(\!\frac{1}{(\bj)^{10/3}}\!\right) \, .
\end{equation}

\begin{figure}[h!]
    \centering

    \subfigure[$n=3/2$]{
        \includegraphics[width=0.48\textwidth]{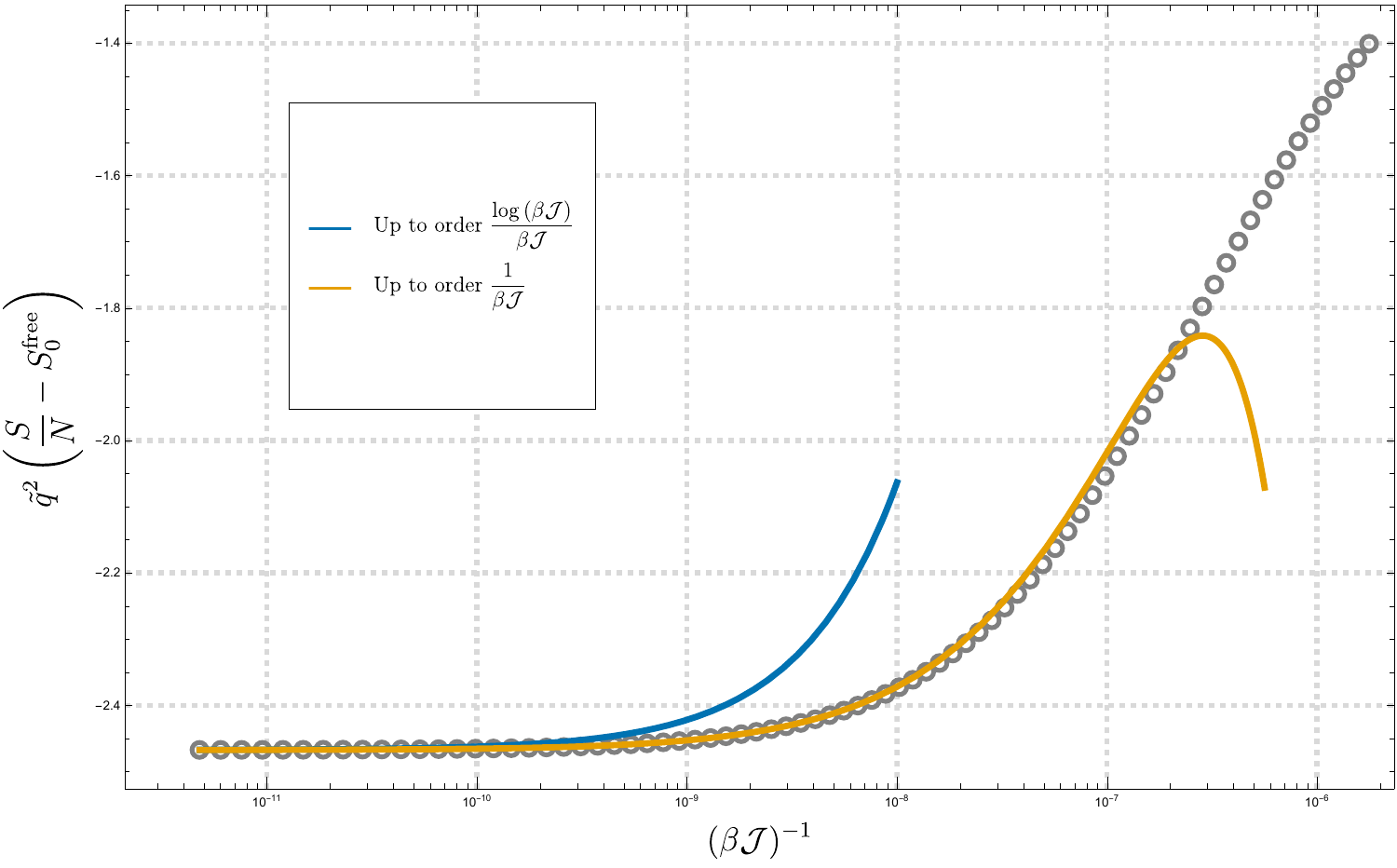}
        \label{fig:3_over_2_corrections}
    }
    \hfill
    \subfigure[$n=4/3$]{
        \includegraphics[width=0.48\textwidth]{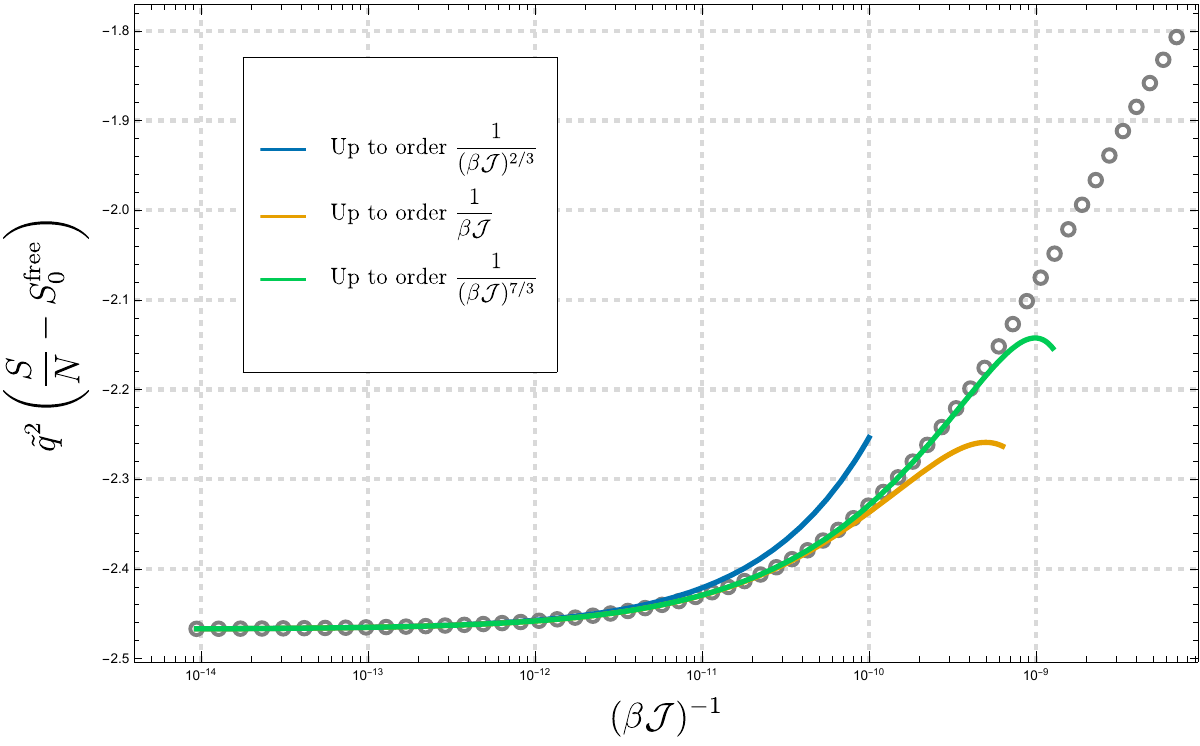}
        \label{fig:4_over_3_corrections}
    }

    \caption{Entropy as a function of temperature for two different values of $n$. The circles are numerical values of $\tilde{q}^2 \left(\frac{S}{N}-S_0^{\text{free}} \right)$ with $s=0.01$. The solid lines correspond to the low-temperature analytical expansion, truncated at different orders.}
    \label{fig:entropy corrections}
\end{figure}

When inserted into the entropy formula \eqref{Entropy}, this gives
\begin{equation}
    \label{4/3 entropy}
    \frac{S}{N}=S_0^{\text{free}}-\frac{\pi^2}{4\tilde{q}^2}+\frac{\tilde{S}_1\!\left(n=\frac{4}{3},s\right)}{\tilde{q}^2}\frac{1}{(\bj)^{2/3}}+\frac{S_1\!\left(n=\frac{4}{3},s\right)}{\tilde{q}^2}\frac{1}{\bj}+\mathcal{O}\!\left(\!\frac{1}{(\bj)^{4/3}}\!\right) \, ,
\end{equation}
where
\begin{equation}
    \label{S1 tilde 4/3}
    \tilde{S}_1 \!\left(\!n=\frac{4}{3},s \!\right)=\frac{15 \pi ^{13/6} \Gamma \left(-\frac{5}{6}\right)}{64  \Gamma \left(-\frac{1}{3}\right) s^{8/3}} \, , \quad S_1\!\left(\!n=\frac{4}{3},s\! \right)=\frac{\pi ^2 \left(8 s^2-9\right) \sqrt{9+16 s^2}}{32s^4} \, .
\end{equation}
Note that $\tilde S_1(n=4/3, s)$ is consistent with \eqref{eq: tilde S1}. This low-temperature expansion is compared to the numerical evaluation of the entropy in figure \ref{fig:4_over_3_corrections}, showing excellent agreement.

\begin{figure}[!htbp]
    \centering

    \subfigure[]{
        \includegraphics[width=0.48\textwidth]{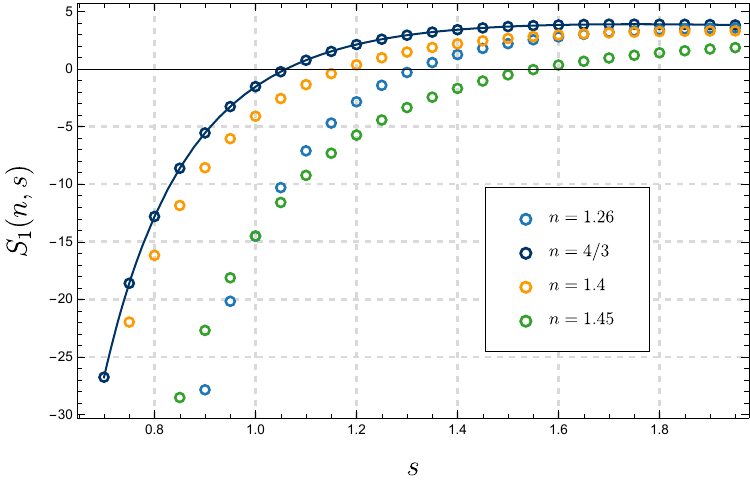}
        \label{fig: numerical subleading linear}
    }
    \hfill
    \subfigure[]{
        \includegraphics[width=0.48\textwidth]{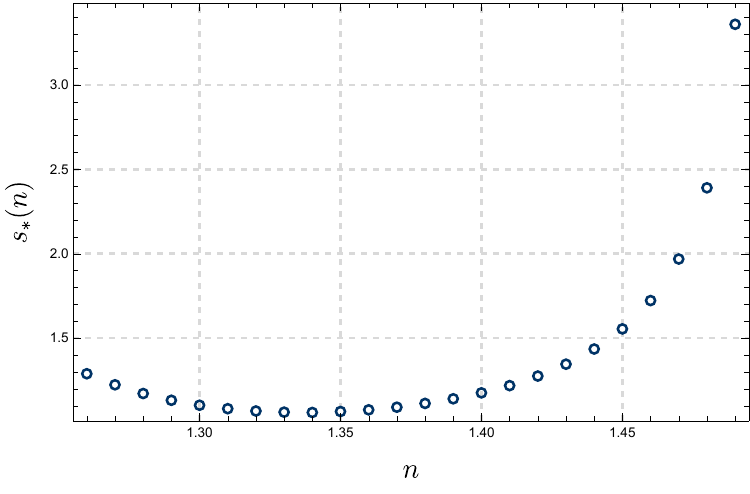}
        \label{fig: roots of linear coef}
    }

    \caption{(a) The circles are numerical values of $S_1(n,s)$ as a function of $s$, while the solid curve is the analytical expression from \eqref{S1 tilde 4/3} in the case $n=4/3$. For each value of $n$, $S_1(n,s)$ vanishes and changes sign at some value of $s$, denoted $s_*(n)$; (b) The numerical value of $s_{*}(n)$ as a function of $n$.} 
    \label{fig:subleading linear vanishing}
\end{figure}

Interestingly, the coefficient of the linear-in-temperature term changes sign as we vary $s$, and \textit{vanishes} at $s_*=\frac{3}{2\sqrt{2}}$. We verify this in figure \ref{fig: numerical subleading linear}. Moreover, in the same figure, we provide numerical evidence that this is generally the case for all anomalous deformations when $5/4 < n < 3/2$. In such cases, there exists $s_*(n)$ for which the linear-in-temperature term vanishes. We compute these values using the bisection method and plot them in figure \ref{fig: roots of linear coef}.\footnote{Clearly, this data is not smoothly connected to the exact $n=3/2$ result, where we get $s_* (n=3/2) \approx 0.7126$, see figure \ref{fig:n=1.5_linear}. This is justified by the presence of additional logarithmic terms in the $n=3/2$ case.} 

Note that in figure \ref{fig: roots of linear coef}, all values of $s_{*}$ exceed 1. As we show in section \ref{sec: intermedIR}, in order to have an intermediate IR, it is necessary to have $s\ll1$. This implies that, for the cases in which we have both an intermediate IR regime and an anomalous scaling in the deep IR, the linear-in-temperature contribution in the deep IR will always be strictly negative.

\subsection{Finite $q,\tilde{q}$} \label{sec: Finite q} 

So far, we have shown that in the large $q,\tilde{q}$ limit of the deformed SYK model with fixed $1<n=q/\tilde q<3/2$, there exists a deep infrared regime in which the entropy scales with temperature in an anomalous way, overcoming the standard linear-in-temperature behaviour in Schwarzian-dominated theories.

Here we explore the question of whether this anomalous scaling survives at finite $q, \tilde q$.\footnote{It is also of interest to determine whether this anomalous scaling remains valid at finite $N$. In this case, the only available techniques are those of exact (or approximate) diagonalisation. This is beyond the scope of the present work, but we hope to report on finite $N$ results in future communication.} To answer this question, we numerically solve the Schwinger-Dyson equations \eqref{Schwinger-Dyson} for some values of $q,\tilde{q}$ such that $1 < q/\tilde{q} < 3/2$. 
The numerical algorithm is based on the recursive application of the fast Fourier transform, as first developed in \cite{Maldacena:2016hyu}. We apply the numerical implementation used in \cite{Anninos:2022qgy, Sheorey:2024bvt}.\footnote{The code used is accessible to the reader at: \url{https://github.com/sameersheorey}.} The numerical procedure at finite $q, \tilde q$ is much more computationally demanding, especially at low temperatures. In fact, we can reliably reach inverse temperatures only of the order $\bj \approx 10^{2}$. In order to reach the deep infrared regime, we therefore take a value of $s$ which is not very small. 

In what follows, we use $s=1$ which, as we will see in section \ref{sec: intermedIR}, is not small enough to allow for a parametrically large intermediate infrared regime. Nevertheless, this is enough to characterise the leading scaling behaviour of the deep infrared thermodynamics.

For fixed $q,\tilde q$ and $s$, the algorithm converges to some saddle-point solution $(G_{*},\Sigma_{*})$ from which we can compute the on-shell action. Repeating this process at different temperatures allows us to compute the thermal entropy using \eqref{thermodynamics}.

\begin{figure}[H]
    \centering
    \includegraphics[width=0.55\textwidth]{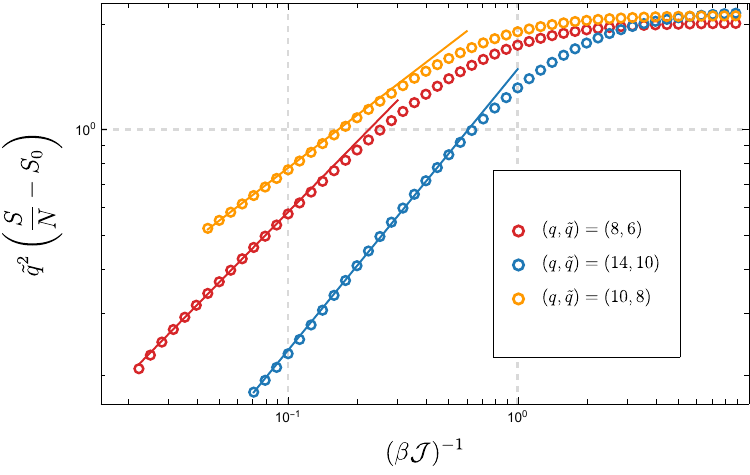}
    \caption{Entropy as a function of temperature for finite values of $q,\tilde q$ and $s=1$. The circles are numerical values of $\frac{S}{N}-S_0$ as a function of $(\bj)^{-1}$. The values of $q,\tilde{q}$ are chosen such that $1<n=q/\tilde{q}<3/2$. As a guide, we show solid lines with anomalous scaling $(\bj)^{2-2n}$.}    \label{fig: finite q}
\end{figure}

As we are working in the saddle-point approximation at large $N$, we still expect the system to have a large zero-temperature entropy at finite $q,\tilde q$. Given the results in section \ref{sec: zero-temp entropy}, an informed guess for this zero-temperature entropy is that it will take the same value as that of a single SYK with $\tilde q$ interactions \cite{kitaev_vid},
\begin{equation}
    \label{finite q zero-temp entropy}
    S_0 \equiv \left. \frac{S}{N} \right|_{\bj \to \infty} = S_0^{\text{free}} - \int_{0}^{1/\tilde{q}}dx \, \pi \left(\frac{1}{2}-x\right)  \tan(\pi x)  \, .
\end{equation}
We have checked that this expression gives the right zero-temperature entropy for our deformed models to good approximation. Subtracting this constant contribution, we can analyse the leading contribution to the entropy away from zero temperature. This is shown in figure \ref{fig: finite q}, for three different cases of finite $q, \tilde q$, with $n$ in the anomalous range. For $n>3/2$, similar numerical results can be found in section 4.2 of \cite{Anninos:2022qgy}, where it is shown that the linear-in-temperature term dominates at low temperatures. 

Here, instead, we find that the entropy in the infrared regime exhibits the same temperature-dependence as in the large $q,\tilde{q}$ limit. In particular, as shown in figure \ref{fig: finite q}, there is a good match between the expected anomalous $(\bj)^{2-2n}$ scaling and the numerical computations. 

\section{Intermediate infrared thermodynamics}\label{sec: intermedIR}
For small $s$, the $n=2$ model analytically revealed the existence of an intermediate infrared regime, where the entropy scales linearly with temperature  (with a coefficient that does not depend on $s$); see section \ref{sec: deformedSYK}. In this regime, the temperature is still low, but not the lowest scale in the problem. In \cite{Anninos:2022qgy}, this intermediate regime was also found numerically for $n\geq2$. Here we extend this result to characterise the intermediate infrared near-fixed point analytically for all values of $n>1$. In particular, we also generalise the regime of validity of this intermediate IR phase. 

\begin{figure}[!htbp]
    \centering
    \includegraphics[width=0.55\textwidth]{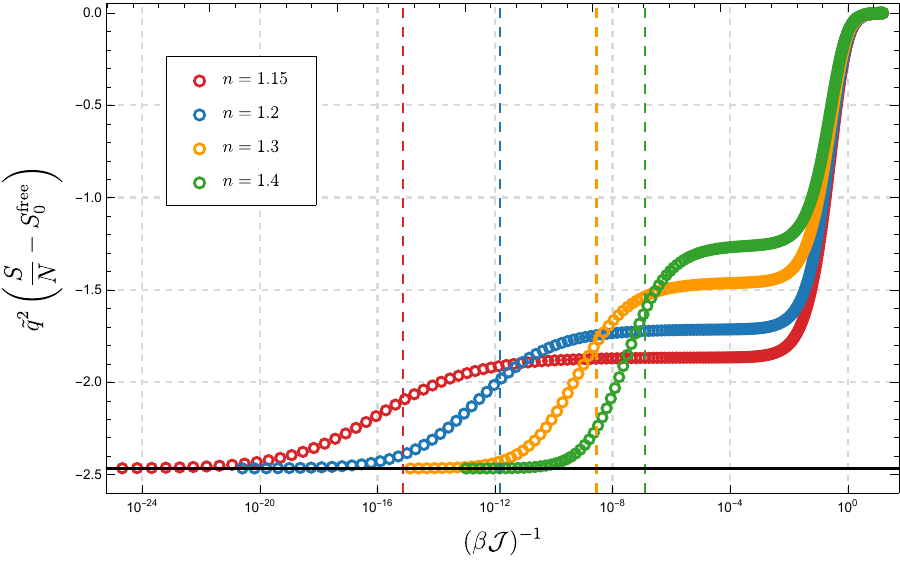}
    \caption{Entropy as a function of temperature across the full thermal RG flow, for a fixed value of $s=10^{-2}$ and different values of $n$. The circles are numerical values of $\tilde{q}^2 \left(\frac{S}{N}-S_0^{\text{free}} \right)$ as a function of $(\bj)^{-1}$. The vertical dashed coloured lines are located at $s^{\frac{n}{n-1}}$, which theoretically separate the intermediate from the deep IR, as in \eqref{intermed inequality}. The horizontal black line at $-\frac{\pi^2}{4}$, corresponds to the zero-temperature entropy in \eqref{zero-temp entropy}.}    \label{fig: full flow n=1.3 s=0.01}
\end{figure}

We start by numerically observing the existence of an intermediate scaling of the entropy at low temperatures. The new numerical procedure described in section \ref{sec: deformedSYK} gives us enough precision at low temperatures to track the full RG flow, even for small $s$. For a fixed $s=0.01$, we show in figure \ref{fig: full flow n=1.3 s=0.01} the full behaviour of the entropy as a function of temperature for different values of $n$. We observe the existence of two different scaling regimes (which appear as plateaus in the plot) at low temperatures. The one appearing at the lowest temperatures is the deep IR, carefully analysed in section \ref{sec: DeepIR}. The other plateau, observed over a separate range of temperatures (still in the infrared), is the intermediate IR regime we characterise next.

Note that the range of validity of this intermediate fixed point depends heavily on $n$. We are able to analytically track this by looking at approximate solutions to \eqref{ODE} in the regime where
\begin{equation}
    \label{intermed inequality}
    1 \ll \beta \mathcal{J} \ll s^{-\frac{n}{n-1}} \, .
\end{equation}
The procedure is exactly analogous to the one we performed in section \ref{sec: DeepIR} for the deep infrared, except that in this case we fix the quantity $\beta \mathcal{J}s^{\frac{2n}{3n-2}}$ to be of order one before performing the perturbative expansion.\footnote{For fixed $s$, the deep IR region emerges at lower temperatures as $n\to1$. In the case considered in figure 1 in \cite{Anninos:2022qgy}, the lowest relative coupling was fixed to $s^2=0.01$, and the temperatures considered were $\bj \sim 10^3$, which is not in the deep IR region for small values of $n$, see figure \ref{fig: full flow n=1.3 s=0.01}.}

We expand $g(t)$ in this regime for any $n>1$, finding that non-trivial equations of motion arise if
\begin{equation}
\label{Intermed g}
    e^{g(t)/n}=\frac{u_0(t)}{(\beta \mathcal{J})^{2/n}}+\frac{u_1(t)}{(\beta \mathcal{J})^{1+\frac{2}{n}}}+ \mathcal{O}\!\left(\!\frac{1}{(\bj)^{2+\frac{2}{n}}}\!\right) \, ,
\end{equation}
with the coefficients solving 
\begin{equation}
    \label{perturbative ODEs intermed}
    \begin{cases}
            \partial_t^2u_0(t)=\frac{(\partial_tu_0(t))^2}{u_0(t)}+\frac{2}{n}u_0(t)^{1+n} \, , \\
            \partial_{t}^2u_1(t)=\frac{2\partial_tu_0(t)}{u_0(t)}\partial_{t}u_1(t)+\left(\frac{2n+4}{n}u_0(t)^{n}-\frac{\partial_{t}^2u_0(t)}{u_0(t)} \right)u_1(t)+2\left(\bj s^{\frac{2n}{3n-2}}\right)^{\frac{3n-2}{n}}u_0(t)^2 \, .
    \end{cases}
\end{equation}
As before, to solve \eqref{perturbative ODEs intermed}, we still need to impose appropriate boundary conditions. Unlike in the case of the deep infrared, here we can completely determine the integration constants for both $u_0(t)$ and $u_1(t)$ analytically, see appendix \ref{sec: intermed appendix}. We find that $u_0(t)$ is given by 
\begin{equation}
\label{eq: u0}
        u_0(t) = \left( \frac{\pi}{\sin (\pi t)}\right)^{2/n} \, .
\end{equation}
As for $u_1(t)$, the solution is analytical but cumbersome, see \eqref{intermed u1 at all t}. To compute the entropy, we only need the value of the function at $t=1/2$,
\begin{equation}
\label{Intermed g coef}
        u_1\!\left( t= \tfrac{1}{2} \right) = \left(\frac{(n-2)  \Gamma \left(\frac{1}{2}-\frac{1}{n}\right)}{\sqrt{\pi} \ \Gamma \left(\frac{n-1}{n}\right)}-n \right) \pi^{\frac{4}{n}-2} \left(\beta  \mathcal{J} s^{\frac{2 n}{3 n-2}}\right)^{3-\frac{2}{n}}-\frac{4 \pi ^{2/n} }{n} \, .
\end{equation}

\begin{figure}[!htbp]
    \centering

    \subfigure[$n=1.2,s=10^{-2}$.]{
        \includegraphics[width=0.48\textwidth]{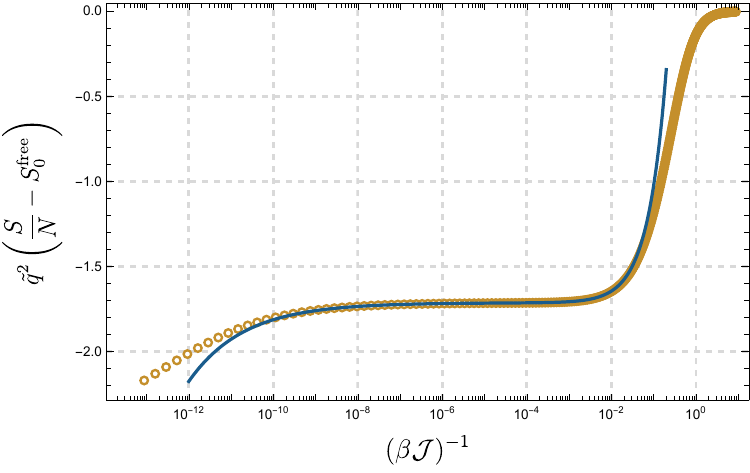}
        \label{fig:intermed plot n=1.2}
    }
    \hfill
    \subfigure[$n=1.7, s=10^{-3}$.]{
        \includegraphics[width=0.48\textwidth]{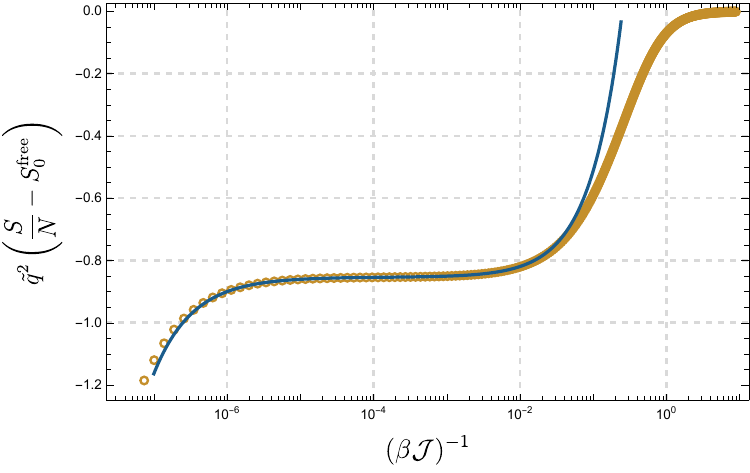}
        \label{fig:intermed plot n=1.7}
    }

    \caption{Entropy as a function of temperature in the intermediate infrared regime. The circles are numerical values of $\tilde{q}^2\!\left(\frac{S}{N}-S_0^{\text{free}}\right)$ as a function of $(\bj)^{-1}$. The solid curves correspond to the analytical expression from \eqref{Intermed Entropy}.}
    \label{fig: intermed plots}
\end{figure}

 Plugging the results into \eqref{Entropy}, we find the entropy in the intermediate infrared regime, 
\begin{equation}
\label{Intermed Entropy}
    \frac{S}{N}=\left[S_0^{\text{free}}-\frac{\pi^2}{4q^2}+\frac{\pi^2}{q^2}\frac{1}{\beta \mathcal{J}}\right]+\left(\frac{2}{n}-1 \right)\frac{\pi^{\frac{2}{n}-\frac{1}{2}}\Gamma\!\left(\frac{1}{2}-\frac{1}{n}\right)n^2}{4\Gamma\!\left(1-\frac{1}{n}\right)q^2}s^2(\beta \mathcal{J})^{2-\frac{2}{n}}+ \mathcal{O}\!\left(\!\frac{1}{(\bj)^2}\!\right) \,.
\end{equation}
The terms inside the square brackets are precisely those of the infrared entropy of a single $n=1$ SYK model with $q$-fermion interactions and $s=0$, see \eqref{n=1 entropy}. In particular, it does not depend on $n$ or $s$. The term proportional to $(\beta \mathcal{J})^{2-\frac{2}{n}}$ is the first correction from the infrared near-fixed point. This contribution matches the conformal perturbation theory result, where the conformal theory is perturbed by a relevant operator of dimension $\Delta=1/n$, as shown in appendix \ref{sec: appendix on conformal perturbation theory}, see \eqref{entropy CPT}. In order to match both results, the coupling constant in the perturbed theory needs to be identified with
\begin{equation}
\label{Coupling}
    g_{\Delta=1/n}^2=\frac{n^2s^2\mathcal{J}^2}{2q^2} \, ,
\end{equation}
which is indeed small if the intermediate infrared regime exists. Note that this result was numerically found in \cite{Anninos:2022qgy}, focusing on the regime $n\geq2$. Here, we analytically reproduced that result, while extending it to all values of $n>1$. In figure \ref{fig: intermed plots}, we show that the analytical result is indeed a good match with the numerical evaluation of the entropy in the intermediate infrared regime. Note that while we have implemented this deformation at the level of the Hamiltonian, the above results confirm that it can indeed be thought of as a small deviation away from the infrared near-fixed point of the single SYK model $H_{q}$.

\section{Out-of-time-ordered correlators} \label{sec: OTOCs}
So far, we have analysed equilibrium probes of the deformed SYK theory \eqref{Deformed Hamiltonian} using Euclidean methods. We showed that for $1<n<3/2$, the large $N$ entropy of the system in the deep infrared scales with an anomalous, stronger-than-linear power of the temperature. In this section, we study whether this feature affects dynamical probes of the system. In particular, we focus on the Lorentzian out-of-time-ordered correlation function,
\begin{equation}
    \label{OTOC}
    \text{OTOC}(t) \equiv \frac{1}{N^2} \sum_{i,j=1}^{N} \text{Tr} \left(\rho^{1/4} \psi_i(t)\rho^{1/4}\psi_j(0)\rho^{1/4}\psi_i(t)\rho^{1/4}\psi_j(0) \right) \, ,
\end{equation}
where $t$ is Lorentzian time and $\rho$ is the thermal density matrix at inverse temperature $\beta$. In the large $N$ limit, chaotic quantum systems with a large separation between dissipation and scrambling are expected to have OTOCs that behave generically as
\begin{equation}
\label{OTOC form}
    \text{OTOC}(t)=f_0-\frac{f_1}{N}\exp(\lambda t) + \cdots \, ,
\end{equation}
where $f_{0,1}$ are order-one positive constants. The expression should be understood as valid from times of order the inverse temperature $\beta$ to times of order the scrambling time $t_\text{sc} \equiv \lambda^{-1} \log N$. Here, we are further assuming that the inverse temperature does not scale with $N$. The single SYK model is known to be maximally chaotic in the infrared regime \cite{kitaev_vid}; in other words, its Lyapunov exponent $\lambda$ saturates the bound on chaos, \cite{Maldacena:2015waa}
\begin{equation}
    \label{Bound on Chaos}
    \lambda \leq \frac{2\pi}{\beta} \, ,
\end{equation}
in the vicinity of the near-fixed point where the Schwarzian action dominates.

As in the previous sections, we focus here on the large $N$ limit and take the large $q,\tilde{q}$ limit with $q/\tilde{q}=n$ fixed. The Lyapunov exponent for the $n=2$ model was studied in \cite{Jiang:2019pam}, where it was observed that, for small $s$, there exist two regions of near-maximal chaos that correspond to the intermediate and deep infrared regimes of the thermal RG flow. In fact, it was analytically shown that 
\begin{equation}
\label{n=2 lyapunov perturbatively}
  \lambda^{(n=2)} =   \begin{cases}
        \frac{2\pi}{\beta} \left(1-\frac{2}{\bj} -\left(\frac{1}{2}-\frac{4}{\pi^2}\right)\bj s^2+ \mathcal{O}\!\left(\!\frac{1}{(\bj)^2}\!\right) \right) \, , \, & \text{for} \, 1\ll \frac{1}{s^2} \ll \bj  \,, \\
        \frac{2\pi}{\beta} \left(1-\frac{\sqrt{1+4s^2}}{\bj s^2} + \mathcal{O}\!\left(\!\frac{1}{(\bj)^2}\!\right) \right) \, , \, & \text{for} \, 1\ll \bj \ll \frac{1}{s^2} \,.
    \end{cases}
\end{equation}
We expect similar qualitative behaviour for $n>3/2$, see numerical evidence in \cite{Chapman:2024pdw}. Here, we focus on $n<3/2$, where the entropy scales anomalously with the temperature in the deep infrared.

\begin{figure}[H]
    \centering

    \subfigure[$n=1.3$]{
        \includegraphics[width=0.48\textwidth]{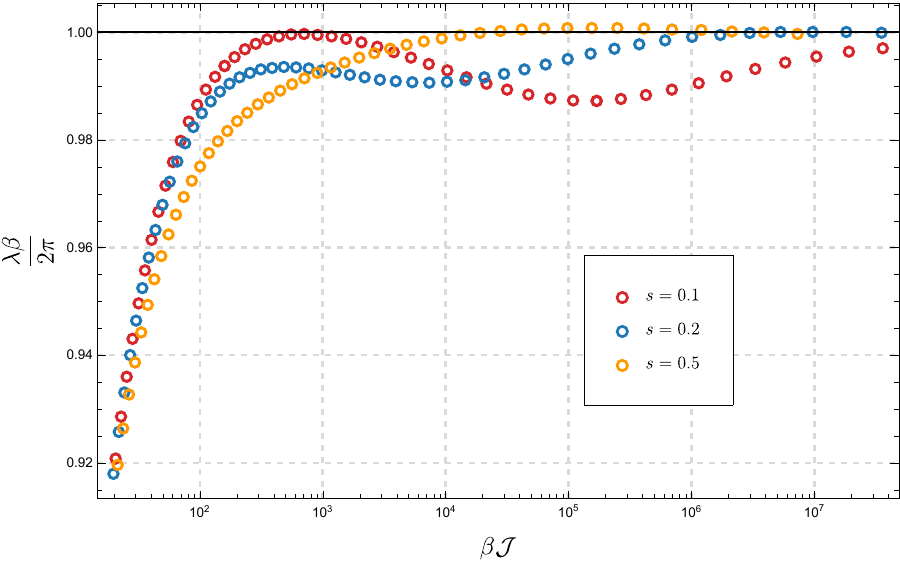}
        \label{fig:lyapunov_n=1.3}
    }
    \hfill
    \subfigure[$n=1.4$]{
        \includegraphics[width=0.48\textwidth]{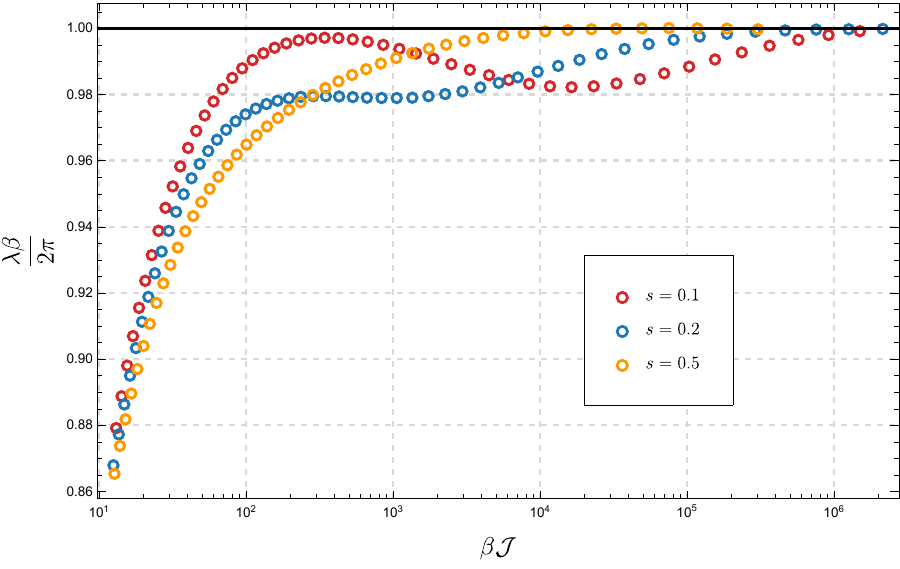}
        \label{fig:lyapunov_n=1.4}
    }

    \caption{The Lyapunov exponent as a function of the inverse temperature for different values of $s$. The circles correspond to numerical values of $\frac{\lambda \beta}{2\pi}$ as a function of $\bj$. The black horizontal line corresponds to maximal chaos. The system is always maximally chaotic at large $\bj$, and another near-maximal chaos region emerges at intermediate low temperatures when $s$ is small.}
    \label{fig:lyapunov}
\end{figure}

In order to compute the Lyapunov exponent at large $N$, we follow the procedure described in \cite{Jiang:2019pam}. Given the action \eqref{large-q action} at large $q,\tilde{q}$, we expand around the classical saddle,
\begin{equation}
    \label{Expand around saddle}
    g(\tau_1,\tau_2)=g_c(\tau_1-\tau_2)+h(\tau_1,\tau_2) \, ,
\end{equation}
and evaluate the analytically continued action to quadratic order in $h$. This is given by 
\begin{equation}
    \label{action around saddle}
    \frac{I}{N} = I_{\text{on-shell}} - \frac{1}{32q^2} \int dt_{+} \, d\sigma \left[(\partial_{t_{+}} h)^2 - (\partial_{\sigma}h)^2 + 2\mathcal{J}^2(s^2 e^{g_c/n}+e^{g_c})h^2\right]  +\mathcal{O}(h^3) \, ,
\end{equation}
where $t_{+} \equiv -i(\tau_1 + \tau_2)$ and $\sigma \equiv -i(\tau_1-\tau_2)$. Then, the thermal correlator,
\begin{equation}
    \label{thermal correlator}
    K(t_{+},\sigma)= \langle h(t_{+},\sigma)h(0,0) \rangle_{\beta} \, ,
\end{equation}
being the inverse of the quadratic part of the action, solves
\begin{equation}
    \label{green}
    \left[-\partial_{t_{+}}^2 + \partial_{\sigma}^2 +2\mathcal{J}^2 (s^2 e^{g_c/n}+e^{g_c})\right]K(t_{+},\sigma) = \frac{16iq^2}{N} \delta(t_{+}) \delta(\sigma) \, .
\end{equation}
Inserting the late-time ($t_{+} \gg 1$) ansatz $K(t_{+},\sigma) = e^{\lambda t_{+}/2}f_{\lambda}(\sigma)$, where $\lambda$ is the Lyapunov exponent, we find the ordinary differential equation
\begin{equation}
    \label{4pt ODE in terms of sigma}
    \left(\frac{\lambda^2}{4}-\partial_{\sigma}^2 \right) f_{\lambda}(\sigma) = 2\mathcal{J}^2 \left(s^2 e^{g_c(\sigma)/n}+e^{g_c(\sigma)} \right) f_{\lambda}(\sigma) \, .
\end{equation}
This equation requires knowledge of the analytically continued two-point function $g_{c}(\sigma)$. However, in the same way the thermodynamic quantities depend only on the value $g_m=g(t=1/2)$, we may simplify equation \eqref{4pt ODE in terms of sigma} as follows. Applying the change of variable $\sigma(g_c)$, which follows from the analytic continuation of \eqref{gstar integral},
\begin{equation}
    \label{sigma in terms of g}
    \sigma = \frac{1}{2\mathcal{J}} \int_{g_c}^{g_m} \frac{dx}{\sqrt{W(g_m)-W(x)}} \, .
\end{equation}
Then, \eqref{4pt ODE in terms of sigma} becomes
\begin{equation}
    \label{final lyapunov ODE}
    \left(\frac{\lambda^2}{4}f_{\lambda}(g)-4\mathcal{J}^2 \sqrt{W(g_m)-W(g)}\partial_{g} \left(\sqrt{W(g_m)-W(g)} f_{\lambda}'(g) \right) \right)  = 2\mathcal{J}^2(s^2 e^{g/n}+e^{g}) f_{\lambda}(g) \, ,
\end{equation}
where we removed the subscript ``$c$'' for simplicity of notation. This equation is further supplemented by the condition $f_{\lambda}'(g_m)=0$. To determine the Lyapunov exponent, we implement the shooting method to numerically solve \eqref{final lyapunov ODE} using the Runge-Kutta algorithm to find $\lambda$ so that the solution is normalisable. In figure \ref{fig:lyapunov} we plot $\lambda$ as a function of inverse temperature $(\bj)^{-1}$ for some values of $s$ and two anomalous values of $n$.

Evidently, the system approaches maximal chaos (to numerical precision) in the deep IR and, for small $s$, near-maximal chaos in the intermediate regime.\footnote{In order to look for normalisable solutions numerically, we manually choose the smallest boundary value $g=g_{b}$ which is `far enough' so that the solution to \eqref{final lyapunov ODE} falls off and reaches values of the order $10^{-7}$ at $g_{b}$. The final solution may be sensitive to this number $g_{b}$. For each value of $\b$, we find that the value of $g_{b}$ which gives results closest to the analytical predictions in \eqref{ly} and \eqref{intermediate lyapunov} is the minimum one for which the solution appears normalisable. For the temperatures considered here, these values range from $-16$ to $-40$.
}
Note that in the region in between the two near-fixed points the Lyapunov exponent decreases and then increases again to reach maximal chaos in the deep IR. It would be interesting to further characterise this non-monotonic behaviour.

These deformed SYK models with $1<n<3/2$ thus provide examples of theories with maximal chaos, that are not dominated by the Schwarzian at low temperatures. In what follows, we verify this claim analytically and compute the leading corrections away from maximal chaos.

To do this, we first observe that equations \eqref{4pt ODE in terms of sigma} and \eqref{final lyapunov ODE} take the form of a Schr\"odinger problem of finding the bound states of a particle in a potential. In general, these equations are hard to solve analytically, even if we had access to exact expressions of $g_c(\sigma)$ or $g_m$ at finite temperature. However, we may analyse this equation perturbatively around each near-fixed point where we have the expression for $g_c(\sigma)$; this is how the expansions in \eqref{n=2 lyapunov perturbatively} were determined \cite{Jiang:2019pam}.

For $1<n<3/2$, we have determined the two leading terms of $g(t)$ in the deep IR in section \ref{low-temperature entropy}. After analytic continuation, we plug them into \eqref{4pt ODE in terms of sigma} and find a simplified Schr\"odinger problem where the leading term in the Hamiltonian is given by a universal $n$-independent term, whose ground state yields maximal chaos,
\begin{equation}
    \lambda^{(n>1)} \r \frac{2\pi}{\beta} \, .
\end{equation}
This stems from the $n$-independence of the leading term of $g_c(\sigma)$, analytically continued from $g_0(t)$ in \eqref{g0 with unknown, sec3}. The subleading contribution at low temperatures in the Hamiltonian arises due to the anomalous term $h_1(t)$ in \eqref{Solution 1 to perturbative ODEs}. Using the tools of time-independent perturbation theory, we obtain an analytical expression for the first correction away from maximal chaos in the deep IR, which is negative definite, confirming that the deformed SYK models do not violate the maximal chaos bound \cite{Maldacena:2015waa}. More precisely,

\begin{equation}
    \label{ly}
    \lambda^{(1<n<3/2)}=\frac{2\pi}{\beta}\!\left[1\!-\!\left(\!\frac{(2n-1) \Gamma\!\left(\frac{1}{2}-n\right)}{2\pi \Gamma (1-n)}-\frac{(n-1)(2n+1)}{4}\frac{\Gamma(n+1)}{\Gamma(n+\frac{3}{2})}\!\right)\frac{\pi^{2n-\frac{3}{2}}}{n^2s^{2n}(\bj)^{2n-2}}+\mathcal{O}\!\left(\!\frac{1}{\bj}\!\right)\right] \, .
\end{equation}
A similar correction away from maximal chaos was observed in the coupled SYK model of \cite{Milekhin:2021cou}. Note that when $1<n<5/4$, we expect the next correction to scale with $(\bj)^{4-4n}$ instead of $(\bj)^{-1}$. For the particular case of $n=3/2$, we follow the same method and arrive at a simple expression,
\begin{equation}
    \label{lyapunov 3/2}
    \lambda^{(n=3/2)}=\frac{2\pi}{\beta} \left(1-\frac{4}{9s^3}\frac{\log(\bj)}{\bj} + \mathcal{O}\!\left(\frac{1}{\bj}\right) \right) \, ,
\end{equation}
which agrees with the finite part of the expansion of \eqref{ly} around $n=3/2$. 
For $n>3/2$, because we do not have $g_1(t)$ for generic values of $n$, we do not have analytical control over the leading correction to the maximal Lyapunov exponent, but we expect it to be linear in temperature, as in \eqref{n=2 lyapunov perturbatively}. (This is true for the analytical solutions when $n=3,4$, but the expressions are cumbersome.)

The same method can be applied to the intermediate IR where we fix $\bj s^{\frac{2n}{3n-2}}$ to be of order one. In this case, we obtain
\begin{equation}
    \label{intermediate lyapunov}
    \lambda=\frac{2\pi}{\beta}\!\left[1-\!\frac{2}{\bj}\!-\!\left(\!\frac{  (n-1) (n+2) \Gamma\! \left(1+\frac{1}{n}\right)}{\Gamma\!\left(\frac{3}{2}+\frac{1}{n}\right)}-\frac{2n(n-2)\Gamma\! \left(\frac{1}{2}-\frac{1}{n}\right)}{\pi\Gamma\!\left(1-\frac{1}{n}\right)}\!\right)\frac{\pi ^{2/n}s^2(\bj)^{2-\frac{2}{n}}}{4\pi^{3/2}}+\mathcal{O}\!\left(\!\frac{1}{(\bj)^2}\!\right) \right] \, .
\end{equation}
Note that the maximal Lyapunov exponent receives two corrections in the intermediate IR. The first one is the same correction as in a single SYK model and does not depend on $s$. The second correction comes from the relevant deformation and scales as $s^2 (\bj)^{2-\frac{2}{n}}$. Both contributions are negative definite for all $n>1$, thus satisfying the bound on chaos \eqref{Bound on Chaos}.

\begin{figure}[H]
    \centering
    \subfigure[Intermediate IR, $n=1.45$]{
        \includegraphics[width=0.48\textwidth]{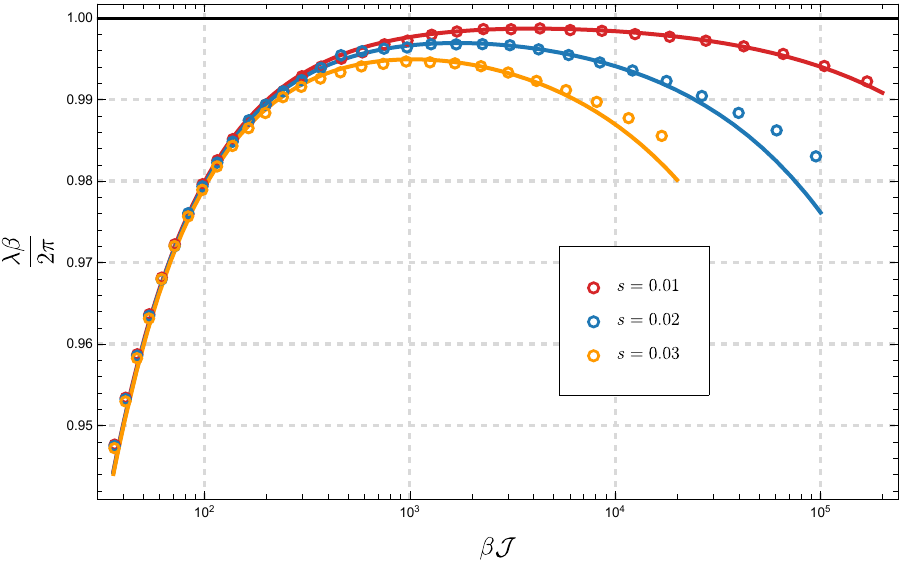}
        \label{fig:lyapunov_intermed}
    }
     \subfigure[Deep IR, $n=1.35$]{
        \includegraphics[width=0.48\textwidth]{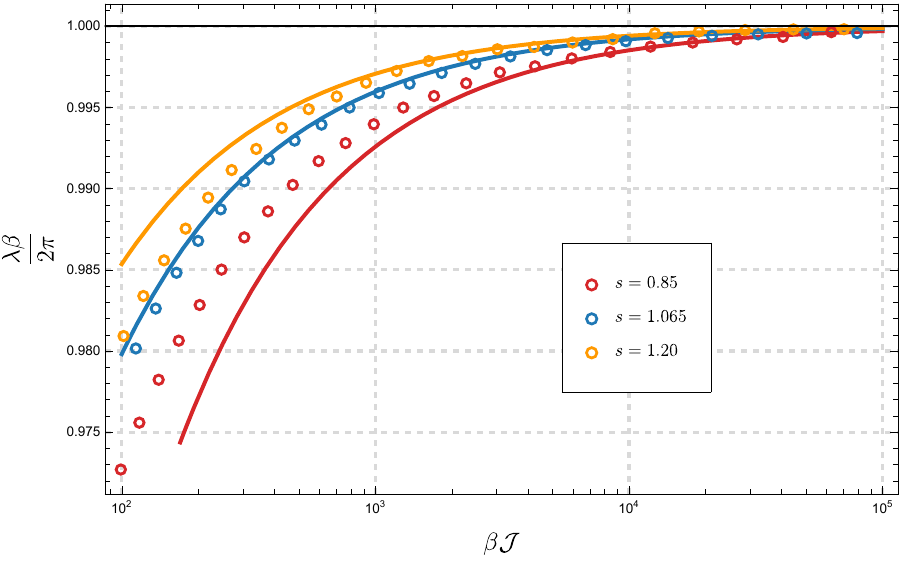}
        \label{fig:lyapunov_deep}
    }
    \hfill
    \caption{The Lyapunov exponent as a function of the inverse temperature for different values of $s$. The circles correspond to numerical values of $\frac{\lambda \beta}{2\pi}$ as a function of $\bj$. The solid, coloured curves correspond to the low-temperature analytical expansions in \eqref{ly} and \eqref{intermediate lyapunov}. The black horizontal line corresponds to maximal chaos.}
    \label{fig:analytical_lyapunov}
\end{figure}

We provide numerical evidence for these analytical results in figure \ref{fig:analytical_lyapunov}. Note that for the value of $n$ chosen in the figure, the next correction in the deep IR is proportional to $(\bj)^{-1}$. As discussed in section \ref{sec: more analytically solvable cases}, this correction (depending on the value of $s$) can have either sign (and even vanish for a particular $s_*$). This can be observed in figure \ref{fig:lyapunov_deep} where the analytical approximation, depending on $s$, can go above or below the full numerical result, and it becomes closer when the value of $s$ is very close to $s_* \approx 1.065$ for $n=1.35$. 

\section{Discussion}
\label{sec:outlook}

In this work, we explored, in a systematic way, curious infrared behaviour that stems from the SYK model. To induce this behaviour, we deformed the SYK Hamiltonian with a $q$-fermion interaction with a disordered operator that has the same form as the Hamiltonian itself, but with a smaller number of fermions $\tilde q$ in the interaction term. We considered the system in the large-$N$ limit and, for most of the paper, focused on the large-$q,\tilde q$ limit with $q/\tilde{q}$ fixed. After taking these limits, the deformed SYK model depends on three dimensionless parameters, $\bj$ (which we interpret as the energy scale of the problem), $s$ (the relative scale between the deformation and the original Hamiltonian), and $n= q/\tilde q>1$ (the ratio between the number of fermions in the two interaction terms).

As a probe for the system, we computed the thermal entropy that stems from the Euclidean partition function after averaging over the disorder. The main result of the paper is that the entropy in the deep infrared regime of the theory admits a rich variety of scalings with respect to the inverse temperature $\bj$, which can be summarised as follows:
\begin{equation}
    \frac{S}{N} = S_0^{\text{free}}-\frac{\pi^2}{4\tilde{q}^2} + \frac{1}{\tilde{q}^2} \begin{cases}
        S_1 (n,s) (\bj)^{-1}  + \cdots\,, \quad & \text{for $n>3/2$ \,,} \\
        \tilde S_1(n, s) (\bj)^{2-2n} + \cdots\,, \quad & \text{for $n<3/2$} \,.
    \end{cases}
    \label{eq: deep IR conclusions}
\end{equation}
Exactly at $n=3/2$, there is a logarithmic enhancement that produces an entropy that scales as $\tfrac{\log \bj}{\bj}$. We computed the function $S_1 (n,s)$ analytically for particular values of $n$, and numerically otherwise. The other function, $\tilde S_1(n, s)$, can be computed analytically, and agrees with the expectation from conformal perturbation theory. Furthermore, when $s$ is small, there is an intermediate regime where the entropy, for all $n>1$, behaves as
\begin{equation}
      \frac{S}{N}=\left[S_0^{\text{free}}-\frac{\pi^2}{4q^2}+\frac{\pi^2}{q^2}\frac{1}{\beta \mathcal{J}}\right]+\left(\frac{2}{n}-1 \right)\frac{\pi^{\frac{2}{n}-\frac{1}{2}}\Gamma\!\left(\frac{1}{2}-\frac{1}{n}\right)n^2}{4\Gamma\!\left(1-\frac{1}{n}\right)q^2}s^2(\beta \mathcal{J})^{2-\frac{2}{n}}+ \mathcal{O}\!\left(\!\frac{1}{(\bj)^2}\!\right) \,.
      \label{eq: int IR conclusions}
\end{equation}
There is a transition between the two regimes that happens at inverse temperatures of order $\beta \mathcal{J} \sim s^{-\frac{n}{n-1}}$. At fixed small $s$, the deep infrared regimes therefore emerges at larger $\bj$ as $n\to 1$.

Of course, these are the simplest relevant deformation one can imagine. More generically, we may include many of these deformations in the Hamiltonian
\begin{equation}
    \label{mult deformations Hamiltonian main text}
    H=H_q+\sum_{n>1} \, s_n H_{q/n} \, \, ,
\end{equation}
where the multiple couplings $s_n$ may be tuned to produce a desired low-temperature behaviour. As we increase $q$, more and more such relevant deformations become available. We develop a particular example of such a theory in the large-$q$ limit in appendix \ref{app: multiple}.  

Given the fact that (a) the deformed SYK models are maximally chaotic in the deep infrared, and (b) the anomalous scalings of the entropy in \eqref{eq: deep IR conclusions} and \eqref{eq: int IR conclusions} are similar to those found in \cite{Horowitz:2022mly}, we end this article by discussing the connections between the thermal RG flows in SYK and the gravitational picture. 

\subsection*{Deformed spacetimes for deformed SYK models}
\label{flow geometrisation}

We consider dilaton-gravity models in two dimensions, consisting of a metric $g_{\mu\nu}$ on a manifold $\mathcal{M}$ and a dilaton field $\phi$ with dilaton potential $U(\phi)$, such that the Euclidean action is given by
\begin{equation}
    \label{dilaton-gravity action}
    S_E = -\frac{1}{2\kappa} \int_{\mathcal{M}}d^2x \sqrt{g} \, \left(\phi R+\ell^{-2}U(\phi) \right) - \frac{1}{\kappa} \int_{\partial \mathcal{M}} du \sqrt{h} \, \phi_{b}K \,,
\end{equation}
where $\kappa$ plays the role of the gravitational constant, $R$ is the Ricci curvature of the metric, $\ell$ is some length scale, $h$ is the induced metric on the boundary $\partial \mathcal{M}$ with extrinsic curvature $K$, and $\phi_b$ is the boundary value of the dilaton. This action assumes Dirichlet boundary conditions where we fix the value of the dilaton and the induced metric on $\partial \mathcal{M}$.\footnote{It may be interesting to consider other boundary conditions, such as those explored in \cite{Banihashemi:2025qqi, Galante:2025tnt}, which allow for macroscopic anomalous behaviour at low temperatures.} In addition, one may include the standard Einstein-Hilbert plus Gibbons-Hawking-York action
\begin{equation}
    \label{Topological action}
    S_0 = - \phi_0 \left(\frac{1}{2\kappa}\int_{\mathcal{M}}d^2x \sqrt{g} \, R +\frac{1}{\kappa} \int_{\partial \mathcal{M}}du \sqrt{h} \, K \right) \, ,
\end{equation}
which in two dimensions is purely topological as a result of the Gauss-Bonnet theorem. The following discussion follows closely \cite{Anninos:2020cwo, Anninos:2022qgy}, with the addition that we now incorporate the anomalous scalings found for $n<3/2$.

The equations of motion that arise from the action \eqref{dilaton-gravity action} admit the following static solution \cite{Cavaglia:1998xj}, 
\begin{equation}
    \label{general metric}
    \frac{ds^2}{\ell^2}=f(r)d\tau^2+\frac{dr^2}{f(r)} \,  , \quad f(r)= \int_{r_h}^{r} d\rho \, U(\phi(\rho)) \, ,  \quad\phi(r)= r\,,
\end{equation}
where $r_h$ is the position of the (Euclidean) horizon. The thermodynamic quantities in the saddle-point approximation are given by \cite{Grumiller:2007ju , Anninos:2017hhn, Witten:2020ert} 
\begin{equation}
    \label{euclidean thermodynamics}
    \beta = \frac{4\pi\ell}{U(\phi(r_h))} \, \, \, \, , \, \, \, \, S=\frac{2\pi \phi_0}{\kappa}+\frac{2\pi}{\kappa}r_h \, . 
\end{equation}
In the microscopic SYK model, we usually compute $S(\beta)$, but it is possible to reconstruct the dilaton-potential simply by inverting it to obtain $\beta(S)$ \cite{Anninos:2020cwo}.

The prototypical example is the single SYK model, whose linear-in-temperature entropy produces a linear dilaton potential $U(\phi) = 2\phi$, characteristic of JT gravity \cite{Maldacena:2016upp}. We now consider the thermal entropy of the deformed SYK models analysed in this work, focusing on the cases in which there exists an intermediate IR regime ($s\ll 1)$.

For $n>3/2$, the situation is qualitatively the same as in the $n=2$ case \cite{Anninos:2020cwo, Anninos:2022qgy}. Both in the deep and in the intermediate IR, there is linear-in-temperature scaling of the entropy, so the dilaton potential interpolates between two regimes where it is linear (but with different slopes). The corresponding saddle-point solution will be a geometry that interpolates between two near-AdS$_2$ spaces with different characteristic lengths.

For $n<3/2$, the situation changes qualitatively. While for $s\ll1$, we retain the near AdS$_2$ boundary that corresponds to the intermediate IR of the deformed SYK model, the deep interior of the geometry is different because of the anomalous scaling of the entropy. Recall that in these cases, $(S/N-S_0)$ scales as $(\bj)^{2-2n}$ which, following \eqref{euclidean thermodynamics}, corresponds to a dilaton potential which scales as\footnote{For $n=3/2$, the associated potential takes the form
$   U(\phi) \propto \frac{-\phi}{W_{-1}(-\phi)} \, ,$
where $W_{-1}$ is a generalized Lambert function, defined as a branch of the inverse of $x \mapsto xe^x$ for $x \gtrsim 0$. For small $\phi$, $U(\phi) \propto -\frac{\phi}{\log(\phi)}$.} 
\begin{equation}
    \label{deep interior potential}
    U(\phi) \propto \phi^{\frac{1}{2(n-1)}}\,,
\end{equation}
for $\phi \ll 1$. The corresponding near-horizon geometry \eqref{general metric} is given by 
\begin{equation}
    \label{anomalous metric}
    f(r) \propto r^{\alpha}-r_h^{\alpha} \, \, \, \, , \, \, \, \, \alpha=\frac{2n-1}{2(n-1)}>2 \, .
\end{equation}
The resulting metric is therefore $\mathcal{C}^2$, but it is not smooth. For example, when $5/4<n<3/2$, the metric is not $\mathcal{C}^3$. Nevertheless, the Ricci scalar $R \propto \alpha(\alpha-1)r^{\alpha - 2}$ and all other curvature invariants remain finite at the horizon. And while some components of the Ricci tensor may diverge in the $(t,r)$ coordinates,
\begin{equation}
    \label{diverging riemann}
    R_{rr} \propto \frac{\alpha(\alpha-1)r^{\alpha-2}}{r^{\alpha}-r_h^{\alpha}} \, ,
\end{equation}
this arises simply because of the choice of coordinates. To verify the absence of null singularities, we work in Lorentzian signature, so we Wick rotate $\tau \r it$. We then define an ingoing null coordinate $v=t+r_*$ where $r_*\equiv\int^{r}\frac{d\rho}{f(\rho)}$ is the usual tortoise coordinate.

In these coordinates, the metric takes the form
\begin{equation}
    \label{metric in ingoing null coordinates}
    ds^2=-f(r)dv^2+2dvdr \,,
\end{equation}
and it is possible to check that all Riemann and Ricci tensor components remain finite. For instance,
\begin{equation}
    \label{finite Riemann tensors}
    R_{vrvr} \propto \alpha(\alpha-1)r^{\alpha-2} \, \, , \, \, R_{vv} \propto \alpha (\alpha-1)r^{2(\alpha-1)}.
\end{equation}

In \cite{Horowitz:2022mly} and related work, the non-smoothness of the horizon persists even after transforming to null coordinates. There is, of course, no reason to expect the microscopic SYK models to display the same behaviour as these deformed higher-dimensional geometries, particularly since the latter are intrinsically non-spherical. Furthermore, in our case, all geometric quantities are computed within the two-dimensional framework. Nevertheless, it would be interesting to explore whether modified theories of gravity, possibly coupled to matter, can reproduce similar low-energy features. It may also be interesting to try to match matter correlation functions in the bulk with those of the deformed SYK model, along the lines of \cite{Anninos:2020cwo, Castro:2025pst}. Alternatively, one may envision constructing deformed SYK models with global symmetries \cite{Anninos:2017cnw, Yoon:2017nig, Gu:2019jub}, which may offer a route towards capturing higher-dimensional, non-spherical aspects of spacetime. We leave these open problems for future work.


\section*{Acknowledgments}
We are grateful to Dionysios Anninos, Shira Chapman, Roberto Emparan,  Ohad Mamroud, Vladimir Narovlansky, and Sameer Sheorey for useful discussions. We are particularly grateful to Sameer Sheorey for allowing us to adapt his code for the numerical simulations used in this project. The work of WAH is funded by the CMA CGM Excellence for Education Scholarship. WAH would like to thank the organizers and participants of the Summer School on The Disordered Universe in Castro Caldelas, Spain, where part of this work was completed. The work of DAG is funded by UKRI Stephen Hawking Fellowship EP/W005530/1 ``Quantum Emergence of an Expanding Universe". DAG is further supported by STFC consolidated grant ST/X000753/1. 

\appendix


\section{Boundary conditions at low temperatures}\label{sec: appendix A}

Here, we discuss the problem of establishing correct boundary conditions for the functions appearing in the low-temperature expansion of the two-point correlator, see sections \ref{sec: DeepIR} and \ref{sec: intermedIR}. The naive low-temperature expansion breaks down close to $t\to0,1$. Instead, we find it more convenient to analyse the solution strictly on the infinite (Euclidean) line, perturbatively at short separations, and from there infer the low-temperature behaviour of the solution at finite but large $\bj$.

\subsection{The deep IR for $n=2$}
\label{appendix A method}

We first discuss the case of $n=2$. As we have the full analytical two-point function \eqref{n=2 solution} in this case, we can use its expansion to test our ideas. We start by recalling the problem with the standard low-temperature expansion. 

Applying the low temperature expansion \eqref{eq: generally g} for $n=2$, we have
\begin{equation}
    \label{g expansion n=2}
    e^{g(t)/2} = \frac{g_0(t)}{(\bj)^2}+\frac{g_1(t)}{(\bj)^3}+\mathcal{O}\!\left(\!\frac{1}{(\bj)^4}\!\right) \, ,
\end{equation}
for some smooth functions $g_0(t),g_1(t)$ satisfying
\begin{equation}
    \label{n=2 perturbative ODEs}
    \begin{cases}
        \partial_t^2g_0(t)=\frac{(\partial_tg_0(t))^2}{g_0(t)}+2s^2g_0(t)^2 \, , \\
        \partial_t^2g_1(t)=\left(\frac{2\partial_{t}g_0(t)}{g_0(t)}\right)\partial_{t}g_1(t)+\left(6s^2g_0(t)-\frac{\partial_{t}^2g_0(t)}{g_0(t)} \right)g_1(t) \, .
    \end{cases}
\end{equation}
The boundary conditions arising from the $t \r 1-t$ symmetry are $\partial_tg_0(t=\frac{1}{2})=\partial_tg_1(t=\frac{1}{2})=0$. Using this information, we can find that $g_0(t)$ is given by
\begin{equation}
    \label{g0 for n=2 with unknown}
    g_0(t)=\frac{c_0^2}{4s^2 \cos^2\left(\frac{c_0}{4}(2t-1) \right)} \,,
\end{equation}
up to some unknown $c_0$ which, without loss of generality, is assumed to be positive. But we still need to impose one more boundary condition.

Consider a complementary expansion in terms of a new variable $x \equiv (\bj t)^{-1}$. The aim is to solve \eqref{ODE} at low temperatures, which we conveniently rewrite here,
\begin{equation}
    \label{resummation ODE n=2}
    \partial_{x}^2Y(x)=\frac{(\partial_{x}Y(x))^2}{Y(x)}-\frac{2}{x}\partial_{x}Y(x)+\frac{2s^2Y(x)^2+Y(x)^3}{x^4} \,,
\end{equation}
in terms of the function $Y(x) \equiv e^{g(x)/2}$. Note that while the differential equation \eqref{resummation ODE n=2} does not explicitly depend on $\bj$, the boundary conditions do. We will solve for the zero-temperature solution $Y_0(x)$, which corresponds to solving the model on the infinite Euclidean line. In this case, the corresponding boundary conditions take the form 
\begin{equation}
    Y_0 (x=0) = 0 \,, \, Y_0(x \to \infty) = 1 \,.
\end{equation}
For $n=2$, remarkably, the solution to $Y_0(x)$ can be fully determined \cite{Anninos:2020cwo},
\begin{equation}
    \label{resummation ODE sol for n=2}
    Y_0(x)=\frac{x^2}{x^2+x \sqrt{4 s^2+1}+s^2} = \frac{x^2}{s^2}-\frac{\sqrt{1+4s^2} \, x^3}{s^4}+\mathcal{O}(x^4) \, .
\end{equation}

But let us assume we do not know the full answer, so we work perturbatively in $x \ll1$. Solving \eqref{ODE} for small $x$, we find that the leading term is necessarily
\begin{equation}
    \label{resummation leading n=2}
    Y_0(x)=\frac{x^2}{s^2}+\mathcal{O}(x^3) \, .
\end{equation}
With this information, we can go back to $g_0(t)$. In terms of $\bj$ and $t$, \eqref{resummation leading n=2} implies that $g_0(t)$ diverges in the $t\to 0$ limit as
\begin{equation}
    \label{leading t=0 divergence g0 for n=2}
    g_0(t)=\frac{1}{s^2t^2}+\cdots \,.
\end{equation}
Matching this with \eqref{g0 for n=2 with unknown} fixes $c_0=(4k+2)\pi$, for some integer $k$. To go to the next order, we may inquire about the subleading terms in \eqref{resummation leading n=2}. For instance, the $x^3$ coefficient would constrain the integration constants in $g_1(t)$ and therefore yield the linear-in-temperature coefficient of the entropy. However, it turns out that the higher-power coefficients of $Y_0(x)$ can only be determined after implementing the boundary condition at $x \r \infty$, which would correspond to fully solving the problem. 

For the case of $n=2$, this is simple to implement. We look at the second term in the small $x$ expansion of \eqref{resummation ODE sol for n=2}, which implies that $g_1(t)$, for small $t$, diverges as
\begin{equation}
    \label{leading t=0 divergence g1 for n=2}
    g_1(t)=-\frac{\sqrt{1+4s^2}}{s^4t^3} \, +\cdots .
\end{equation}
If we plug $g_0(t)$ with $c_0=(4k+2)\pi$ into \eqref{n=2 perturbative ODEs}, we see that $g_1(t)$ only diverges at small $t$ when $k=0$. This directly fixes the remaining integration constant to $c_0 = 2\pi$. For this value of $c_0$, $g_1(t)$ is given by 
\begin{equation}
    \label{g1 for n=2 with unknown}
    g_1(t) = \frac{c_1 (\pi  (2 t-1) \cot (\pi  t)-2)}{\pi \sin^2(\pi t) } \,,
\end{equation}
for some constant $c_1$, which can be fully determined by \eqref{leading t=0 divergence g1 for n=2},
\begin{equation}
    \label{c1 for n=2}
    c_1 = \frac{\pi ^3 \sqrt{1+4 s^2}}{s^4} \, .
\end{equation}
This information is enough to reproduce the low-temperature expansion of the entropy in \eqref{n=2 entropy}. As discussed above, this was only possible because we know the full solution at zero-temperature for $n=2$. More generally for $n>1$, the full solution is not available. Nevertheless, we can still extract some useful information about them in both the deep and the intermediate IR, which is what we do in the following two subsections.

\subsection{The deep IR for arbitrary $n>1$}
\label{appendix A zero-temp entropy}

We can follow the same procedure for more general values of $n$ in the deep IR. The general equation of motion \eqref{ODE} can be now written as
\begin{equation}
    \label{resummation ODE arbitrary n}
    \partial_{x}^2Y(x)=\frac{(\partial_{x}Y(x))^2}{Y(x)}-\frac{2}{x}\partial_{x}Y(x)+\frac{2s^2Y(x)^2+\frac{2}{n}Y(x)^{n+1}}{x^4} \,,
\end{equation}
where $Y(x)=e^{g(x)/n}$. We solve for the zero-temperature solution $Y_0(x)$, so the boundary conditions are the same as in the $n=2$ case,
\begin{equation}
    Y_0 (x=0) = 0 \,, \, Y_0(x \to \infty) = 1 \,.
\end{equation}
In the general case, however, we have not been able to find an analytical solution for $Y_0(x)$. However, the leading term in an $x \ll 1$ expansion is once again fixed,
\begin{equation}
    \label{x squared over s squared}
    Y_0(x)=\frac{x^2}{s^2}+\cdots \, ,
\end{equation}
independently of the boundary condition at $x \r \infty$. Following the same reasoning as in the $n=2$ case, this fixes the result in \eqref{y0 and y0 tilde} and therefore the zero-temperature entropy for all $n>1$.

As with the expansion implemented in section \ref{sec: DeepIR}, the form of the next term differs depending on the value of $n$. If $n>3/2$, the next two terms in the small-$x$ expansion go as $x^{3}$ and $x^{2n}$. Which of the two dominates depends on whether $1<n<3/2$ or $n>3/2$. Either way, it turns out that the coefficient of the $x^{2n}$ term is fixed by the equation \eqref{resummation ODE arbitrary n} without any reference to the boundary condition at $x \r \infty$. It coincides with the coefficient of the $t^{-2n}$ term in the $t \ll 1$ expansion of $h_1(t)$ in \eqref{Solution 1 to perturbative ODEs}, which does not depend on the integration constant $c_2$. 

To fix $c_2$, we note that \eqref{resummation ODE arbitrary n} does not allow a term like $h_1(t)$, which comes with a factor of $(\bj)^{-2n}$, to have a $t^{-3}$ term at $t \ll 1$. Setting the coefficient of the term with $h_1(t)$ to zero fixes $c_2$ to take the value in \eqref{eq: tilde S1}.

If we had access to the solution $Y_0(x)$ at finite values of $x$, each term in the power series $Y_0(x)=\sum_k \alpha_kx^k$ would constrain the solutions to $g_k(t),h_k(t),\ell_k(t)$. For instance, the coefficient of $x^3$, which cannot be fully fixed for generic $n$, would fix the linear-in-temperature coefficient $S_1(n,s)$.

\subsection{The intermediate IR for arbitrary $n>1$} \label{sec: intermed appendix}

We do a similar procedure but now in the intermediate IR. We first fix  $\beta \mathcal{J}s^{\frac{2n}{3n-2}}$ to be order one, such that \eqref{ODE} becomes
\begin{equation}
    \label{intermed resummation ODE}
    \partial_{x}^2Y(x)=\frac{(\partial_{x}Y(x))^2}{Y(x)}-\frac{2}{x}\partial_{x}Y(x)+\frac{2Y(x)^{n+1}}{nx^4}+\left(\frac{\bj s^{\frac{2n}{3n-2}}}{\bj}\right)^{\frac{3n-2}{n}}\frac{2}{x^4}Y(x)^2 \,.
\end{equation}
We solve for $Y_0(x)$, that is, in the limit in which the last term is subleading, and which satisfies the same boundary conditions as before. In this case, we can obtain the full solution for $Y_0(x)$,
\begin{equation}
\label{intermed resummation solution}
    Y_0(x)= \left(\frac{x}{1+x} \right)^{2/n}=x^{2/n} \left(1-\frac{2x}{n} + \mathcal{O}(x^2) \right) \, .
\end{equation}
As in the $n=2$, we can expand the solution for small $x$, and from there obtain the leading corrections to the two-point function in the intermediate IR analytically. This yields the $u_0(t)$ reported in \eqref{eq: u0}. Furthermore, the full solution to $u_1(t)$ in \eqref{perturbative ODEs intermed} is given by
\begin{equation}
    \label{intermed u1 at all t}
    \begin{aligned}
            u_1(t) = \frac{1}{\pi ^{\frac{2 n-4}{n}} \sin ^{\frac{n+2}{n}}(\pi  t)} \left[ -(n-2) \left(\bj s^{\frac{2n}{3n-2}} \right)^{\frac{3 n-2}{n}} \cos ^2(\pi  t) \, _2F_1\left(\frac{1}{2},\frac{n+2}{2 n};\frac{3}{2};\cos ^2(\pi  t)\right)  \right. \\
            \left.  +\frac{(n-2) \Gamma \left(\frac{1}{2}-\frac{1}{n}\right) \left(\sqrt{\pi } \left(\bj s^{\frac{2n}{3n-2}} \right)^{\frac{3 n-2}{n}} (2 \sin (\pi  t)+(\pi -2 \pi  t) \cos (\pi  t))\right)}{2 \pi \,  \Gamma\!\left(\frac{n-1}{n}\right)} \right. \\
            \left. -n \left(\bj s^{\frac{2n}{3n-2}} \right)^{\frac{3 n-2}{n}} \sin ^{\frac{n-2}{n}}(\pi  t) -\frac{4 \pi ^{\frac{2 n-2}{n}} \sin (\pi  t)}{n}+\frac{2 \pi ^{\frac{3 n-2}{n}} (2 t-1) \cos (\pi  t)}{n} \right] \, ,
    \end{aligned}    
\end{equation}
which, at $t=1/2$, gives \eqref{Intermed g coef}.

\section{Conformal perturbation theory}\label{sec: appendix on conformal perturbation theory}

In sections \ref{sec: DeepIR} and \ref{sec: intermedIR}, we found the thermodynamic behaviour in the deep and intermediate IR regimes purely from taking judicious limits of the equations of motion. The entropy \eqref{Intermed Entropy} in particular contains a portion identical to the low-temperature entropy of a pure SYK model, plus an $n$-dependent portion attributed to deforming the model. A similar $n$-dependent expression was found for the coefficient of the anomalous term \eqref{eq: tilde S1} in the deep IR. These contributions are those of an operator of dimension $\Delta=1/n$ with respect to the intermediate near-fixed point, and $\Delta = n$ with respect to the deep near-fixed point. In this appendix, we justify this claim.

Consider the effective action of a theory, at finite inverse temperature $\beta$ with $\tau \sim \tau+\beta$, near a critical point taking the form
\begin{equation}
    \label{Effective action}
    I_{\text{eff}} = I_{\text{CFT}}+g_{\Delta} \int_{0}^{\beta} d\tau \, O_{\Delta}(\tau) \, ,
\end{equation}
for some conformal operator $O_{\Delta}$ of conformal weight $\Delta$ and coupling $g_{\Delta}$. Conformal perturbation theory allows us to study properties of the deformed theory for weak coupling $g_{\Delta}$. Take the free energy for instance,
\begin{equation}
    \label{free energy CPT line 1}
    \beta F = -\log(\mathcal{Z}_{\text{eff}}) \, ,
\end{equation}
where $\mathcal{Z}_{\text{eff}}$ is the partition function. Rewriting it in terms of the Euclidean path integral and applying a perturbative expansion at weak coupling:
\begin{equation}
    \label{CPT}
    \begin{aligned}
        \beta F &= -\log \left(\int [D \psi] e^{-I_{\text{CFT}}} \left(1-g_{\Delta} \int_{0}^{\beta} d\tau \, O_{\Delta}(\tau) + \frac{1}{2}g_{\Delta}^2 \int_{0}^{\beta}\int_{0}^{\beta} d\tau_1 d\tau_2 \, O_{\Delta}(\tau_1) O_{\Delta}(\tau_2) + \mathcal{O}(g_{\Delta}^3) \right) \right) \\
        &= -\log \left( \mathcal{Z}_{\text{CFT}} \left( 1 - g_{\Delta} \int_{0}^{\beta} d\tau \langle O_{\Delta}(\tau) \rangle + \frac{1}{2}g_{\Delta}^2 \int_{0}^{\beta}\int_{0}^{\beta} d\tau_1 d\tau_2 \langle O_{\Delta}(\tau_1) O_{\Delta}(\tau_2) \rangle + \mathcal{O}(g_{\Delta}^3) \right) \right)  \, .
    \end{aligned}
\end{equation}
Here, $[D \psi]$ denotes the path integral measure over generic field content. As the contribution of the one-point function vanishes assuming conformal invariance, we have
\begin{equation}
    \label{final free energy CPT}
    \beta F = \beta F_{\text{CFT}}-\frac{1}{2}g_{\Delta}^2 \int_{0}^{\beta} \int_{0}^{\beta} d\tau_1 d\tau_2 \langle O_{\Delta}(\tau_1) O_{\Delta}(\tau_2) \rangle + \mathcal{O}(g_{\Delta}^3) \, ,
\end{equation}
where $F_{\text{CFT}}$ is the free energy of the fixed point. Furthermore, the two-point function of a conformal operator of dimension $\Delta$ on the thermal circle $S_{\beta}^{1}$ takes, up to a choice of normalization, the general form
\begin{equation}
    \label{2pt function}
    \langle O_{\Delta}(\tau_1) O_{\Delta}(\tau_2) \rangle = N \left(\frac{\pi}{\beta \mathcal{J} \sin\left(\frac{\pi \tau}{\beta} \right)} \right)^{2\Delta} \, ,
\end{equation}
where $\tau \equiv \tau_1 - \tau_2$. Putting these two results together, we find the second-order contribution of the operator $O_{\Delta}$ to the free energy. For $\Delta>\frac{1}{2}$, the integral diverges; we regulate the result by integrating from $\frac{\epsilon}{\mathcal{J}}$ to $\beta - \frac{\epsilon}{\mathcal{J}}$ and extract the finite part of the $\epsilon \r 0$ limit,
\begin{equation}
    \label{free energy CPT}
    \frac{\beta  \delta^2F_{\Delta}}{N} = -\frac{\pi^{2\Delta-\frac{1}{2}}\Gamma \left(\frac{1}{2}-\Delta \right)}{2\Gamma \left(1-\Delta\right)}\frac{g_{\Delta}^2}{\mathcal{J}^2 (\bj)^{2\Delta-2}} \, .
\end{equation}
Following \eqref{thermodynamics}, the corresponding contribution to the entropy  is given by
\begin{equation}
    \label{entropy CPT}
    \frac{\delta^2S_{\Delta}}{N}=(2\Delta-1)\frac{\pi ^{2 \Delta-\frac{1}{2}}\Gamma \left(\frac{1}{2}-\Delta\right)}{2\Gamma(1-\Delta)}\frac{g_{\Delta}^2}{\mathcal{J}^2(\bj)^{2\Delta-2}} \, .
\end{equation}

This is the precise form we find in \eqref{eq: tilde S1} and \eqref{Intermed Entropy} with $\Delta = n$ and $\Delta = 1/n$ respectively. 

As for the contribution $\delta^{p}S_{\Delta}$ of higher $p$-point functions to the entropy, one can argue as above that they should scale as
\begin{equation}
    \label{higher-point functions contribution to entropy}
    \delta^{p}S_{\Delta} \propto \frac{1}{(\bj)^{p(\Delta-1)}} \, ,
\end{equation}
for $p \geq 3$.  This matches the expectation from \eqref{eq: generally g}. More precisely, the contribution of each $h_k(t)$ term to the low-temperature entropy goes as $(\bj)^{2k(1-n)}$. These are then the higher-order contributions of these operators with $\Delta = n$ in the deep IR, with the observation that correlation functions with odd $p$ seem to be subleading at large $N$. This is consistent with the computation of the three-point function in \cite{Gross:2017aos}. Numerical work in \cite{Anninos:2022qgy} at the level of the intermediate IR reveals a similar conclusion. 

Note that when $1<n<5/4$, such contributions to the entropy may also dominate the linear-in-temperature term at low temperatures. For instance, if $9/8<n<7/6$, the low-temperature entropy takes the form 
\begin{equation}
    \label{multiple stronger-than-linear terms}
    \frac{S}{N}=S_0^{\text{free}}-\frac{\pi^2}{4\tilde{q}^2}+\frac{\tilde{S}_1}{\tilde{q}^2}\frac{1}{(\bj)^{2n-2}}+\frac{\tilde{S}_2}{\tilde{q}^2}\frac{1}{(\bj)^{4n-4}}+\frac{\tilde{S}_3}{\tilde{q}^2}\frac{1}{(\bj)^{6n-6}}+\frac{S_1}{\tilde{q}^2}\frac{1}{\bj}+\cdots \, ,
\end{equation}
with some coefficients $\tilde{S}_2$ and $\tilde{S}_3$. More generally, as we turn down $n$ and it crosses a value in the set $\{1+\frac{1}{2k},k\in \mathbb{N}^{*}\}$, there will be an extra stronger-than-linear term in the entropy. Note that the coefficients $\tilde{S}_k$ with $k>1$ are beyond analytical control. 

\subsection{Conformal perturbation theory in the deformed SYK model} \label{sec: cpt}
Here, we analyse the SYK model from the perspective of conformal perturbation theory. In particular, we view the deep IR as a perturbation of the near-fixed point of a single SYK model with Hamiltonian $sH_{\tilde{q}}$. Accordingly, we apply the redefinition $\Sigma \r \Sigma + \partial_{\tau}$ and express our action as $I=I_{\text{CFT}}+I_{\text{UV}}$ where
\begin{equation}
    \label{UV and CFT actions}
    \begin{aligned}
        I_{\text{CFT}} &= -\frac{1}{2}\log{\det(-\Sigma)}+\frac{1}{2}\int_{0}^{\beta}\int_{0}^{\beta} d\tau_1 d\tau_2 \left( \Sigma G-s^2 \mathcal{J}^2 \frac{2^{\tilde{q}-1}}{\tilde{q}^2}G^{\tilde{q}}\right) \, , \\
        I_{\text{UV}} &= \frac{1}{2}\int_{0}^{\beta}\int_{0}^{\beta} d\tau_1 d\tau_2 \left(\delta(\tau_1-\tau_2)\partial_{\tau_2}G-\mathcal{J}^2 \frac{2^{q-1}}{q^2}G^{q} \right) \, .
    \end{aligned}
\end{equation}
The family of saddle solutions at finite temperature to $I_{\text{CFT}}$ is parametrized by diffeomorphisms $\phi$ on the thermal circle of size $\beta$, but are more conveniently written in terms of reparametrisations on the line,
\begin{equation}
    \label{circle to line}
    f(\tau) = \tan \left(\frac{\pi \phi(\tau)}{\beta} \right) \, .
\end{equation}
These saddles are given by
\begin{equation}
    \label{IR saddles}
    G_{f}(\tau_1,\tau_2) = \frac{b}{(s \mathcal{J})^{2\Delta}}\frac{f'(\tau_1)^{\Delta}f'(\tau_2)^{\Delta}}{|f(\tau_1)-f(\tau_2)|^{2\Delta}} \, ,
\end{equation}
where $\Delta \equiv 1/\tilde{q}$ and
\begin{equation}
    \label{b}
    b = \frac{1}{2} \left( \frac{(1-2\Delta)\tan(\pi \Delta)}{\pi \Delta}\right)^{\Delta} \, .
\end{equation}
When we include the Schwarzian action, the reparametrization symmetry is broken and we are left with the unique saddle $\phi(\tau)=\tau$. We now wish to evaluate the contribution of this saddle to $I_{\text{UV}}$. An appropriate way of studying this contribution order by order is to consider the time variables $\tau_+ \equiv \frac{\tau_1+\tau_2}{2}$ and $\tau_{12} \equiv \tau_1-\tau_2$, and expand in powers of $\tau_{12}$ \cite{Rosenhaus:2018dtp}. Moreover, as the integral over $\tau_{12}$ diverges, we regularize the result by integrating over $(\varepsilon/s\mathcal{J},\beta-\varepsilon/s\mathcal{J})$.\footnote{While $\mathcal{J}$ here renders the regulator dimensionless, the $s$ plays no meaningful role. In fact, a more appropriate choice for the regulator would perhaps be $\varepsilon/s^{\frac{n}{n-1}}\mathcal{J}$ so that the linear-in-temperature term in \eqref{non-kinetic expansion} has the expected small $s$ dependence as in \eqref{small s linear coef}. However, this renders the expression in \eqref{kinetic term expansion} unnecessarily cumbersome, so we avoid this choice.}

\noindent
The results of this procedure for $1 < n < 3/2$ are as follows. First, the regulated kinetic term,
\begin{equation}
    \label{kinetic term CPT}
    I_{\text{UV}} \Big|_{\text{kinetic}} = \frac{1}{2} \int_{0}^{\beta} d\tau_{+}\int_{\varepsilon / s\mathcal{J}}^{\beta - \varepsilon/s\mathcal{J}} d\tau_{12}  \, \delta  \!\left(\tau_1-\tau_2-\frac{\varepsilon}{s \mathcal{J}} \right) \partial_{\tau_2}G_f(\tau_1,\tau_2) \, ,
\end{equation}
admits the following expansion in $\varepsilon$,
\begin{equation}
    \label{kinetic term expansion}
    \begin{aligned}
    I_{\text{UV}}\Big|_{\text{kinetic}} = b \Delta \; \Bigg[ \frac{\bj s}{\varepsilon^{1+2\Delta}} &+ \frac{\pi ^2  (\Delta -1)  }{3}\frac{\varepsilon^{1-2\Delta}}{\bj s}+\frac{\pi ^4  (\Delta -2)   (5 \Delta +1) }{90}\frac{\varepsilon^{3-2\Delta}}{(\bj s)^3}   \\
      &+ \frac{\pi ^6  (\Delta -3) \left(35 \Delta ^2+21 \Delta +4\right)}{5670} \frac{\varepsilon^{5-2\Delta}}{(\bj s)^5} + \mathcal{O}(\varepsilon^{7-2\Delta}) \Bigg]  \, , \\
    \end{aligned} 
\end{equation}
where the linear-in-temperature term comes from a Schwarzian action in $f(\tau_{+})$. The only term here that diverges as $\varepsilon \r 0$ is the term proportional to $\bj$, which does not contribute to the entropy; otherwise, all other terms vanish in this limit.

\noindent
As for the non-kinetic term,
\begin{equation}
    \label{non-kinetic CPT}
    I_{\text{UV}} \Big|_{\text{non-kinetic}} = - \frac{2^{q-2}\mathcal{J}^2}{q^2} \int_{0}^{\beta} d\tau_{+}\int_{\varepsilon / s\mathcal{J}}^{\beta - \varepsilon/s\mathcal{J}} d\tau_{12} \; G_f(\tau_1,\tau_2)^q \, ,
\end{equation}
the series expansion in $\varepsilon \ll 1$ is given by
\begin{equation}
    \label{non-kinetic expansion}
    \begin{aligned}
        I_{\text{UV}} \Big|_{\text{non-kinetic}} = &-\frac{2^{q-2}b^q}{q^2}\Bigg[ \frac{1}{2n-1}\frac{\bj}{s \varepsilon^{2n-1}} + \frac{n \pi^2}{3(2n-3)} \frac{\varepsilon^{3-2n}}{\bj s^3} + \frac{n\pi^4(1+5n)}{90(2n-5)}\frac{\varepsilon^{5-2n}}{(\bj)^3 s^5}+\mathcal{O}(\varepsilon^{7-2n}) \Bigg] \\
        &+    \frac{2^{q-2}b^q}{s^{2n}q^2(\bj)^{2n-2}}
    \Bigg[ \frac{1 }{2 n-1 }+\frac{\pi ^2 n  }{3 (2n-3) }+\frac{\pi ^4 n (5 n+1)   }{90 (2n-5)} + \cdots \Bigg]+ \mathcal{O}(\varepsilon) \,.
    \end{aligned}
\end{equation}
As can be seen, there is a single term in the expansion that diverges as $\varepsilon \r 0$, which is also proportional to $\beta$. This term does not contribute to the entropy. Moreover, there is an infinite set of $\varepsilon$-independent terms with the same $(\bj)^{2-2n}$ and $s^{-2n}$ dependence from \eqref{eq: tilde S1}. In fact, the infinite series in the second pair of square brackets also seems to share the same pole structure as the large $q,\tilde{q}$ result in \eqref{eq: tilde S1}.

\section{Exact results} \label{sec: exact appendix}

In this appendix, we add some exact and perturbative solutions to \eqref{ODE}.

\subsection{Expanding around $n=1$}

The first case we consider is solving \eqref{ODE} by expanding about the simplest case, that is $n=1$. Consider then, $n=1+\varepsilon$ with $\varepsilon \ll 1$. The solution $g(t)$ can be written perturbatively in $\varepsilon$ as
\begin{equation}
    \label{1 plus epsilon expansion}
    g(t)=g_{0}(t)+g_1(t)\varepsilon+g_2(t)\varepsilon^2+\mathcal{O}(\varepsilon^3) \, ,
\end{equation}
where 
\begin{equation}
    \label{1+eps ODE}
    \begin{cases}
        \partial_{t}^2g_0(t) &= 2 (\bj)^2\left(1+s^2\right)e^{g_0(t)} \, , \\
        \partial_{t}^2g_1(t) &= 2 (\bj)^2 e^{g_0(t)} \left(\left(1+s^2\right) g_1(t)+s^2 \left( 1- g_0(t) \right)\right) \, , \\
        \partial_{t}^2g_2(t) &= (\bj)^2 e^{g_0(t)} \left(\left(1+s^2\right) \left(g_1(t){}^2+2 g_2(t)\right)+s^2\left(g_0(t){}^2 - 2g_0(t) g_1(t) \right)\right) \, .
    \end{cases}
\end{equation}
The appropriate boundary conditions correspond to fixing $g_i(t=0) = g_i(t=1) = 0$. The solution for $g_0(t)$ is given in \eqref{n=1 solution}. The leading order in $\varepsilon$ was studied in \cite{Anninos:2022qgy}, where it was found that
\begin{equation}
    \label{order eps 1+eps}
    g_1(t)=\frac{s^2}{1+s^2} \log \left( \frac{\cos^2(\nu)}{\cos^2 \left(2\nu \left(t-\frac{1}{2}\right)\right)} \right) = \frac{s^2}{1+s^2} g_0(t)\, .
\end{equation}
With a bit more work, it is possible to solve the next order in $\varepsilon$ to get
\begin{equation}
    \begin{aligned}
g_2(x) = 
\Bigg[ 
     &2 \Big(-\mathrm{Cl}_2(\pi - 2x) \tan x
        - \log^2\!\big(\!\cos \nu \, \sec x\big)
        + \log\!\big(\!\cos \nu \, \sec x\big)
    \Big) \\
    \quad + 
    &\frac{
        \big(\nu - x \cot \nu \, \tan x\big)
        \big(1 - 4 \log(2 \cos \nu)\big)
        + 2\, \mathrm{Cl}_2(\pi - 2\nu)\, (x \tan x + 1)
    }{
        \nu + \cot \nu
    }
\Bigg]
\frac{s^2}{(1 + s^2)^2} \, ,
\end{aligned}
\end{equation}
where $x=2\nu \left(t-\frac{1}{2} \right)$ and $\text{Cl}_2$ is the Clausen function of order 2, defined by
\begin{equation}
    \label{Clausen}
    \text{Cl}_2(x)=-\int_{0}^{x} \log \bigg| \,2\sin\left(\frac{x'}{2} \right) \!\bigg| dx' \, .
\end{equation}
The low-temperature entropy \eqref{Entropy} of this perturbative solution takes the form
\begin{equation}
    \label{entropy of 1+eps}
    \frac{S}{N} = \left(S_0 + \frac{S_1}{\bj} + \mathcal{O}\!\left(\!\frac{1}{(\bj)^2}\!\right) \right) +  \left(S_{\text{log},0}+   \frac{S_{\text{log},1}}{\bj} +  \mathcal{O}\!\left(\!\frac{1}{(\bj)^2}\!\right) \right)\log(\bj) \, ,
\end{equation}
where
\begin{equation}
\label{1+eps entropy coefficients}
\begin{cases}
\displaystyle
S_0 = S_0^{\mathrm{free}}
    - \frac{\pi^2}{4 \tilde{q}^2}
    + \frac{\pi^2}{2 \tilde{q}^2 (1 + s^2)} \, \varepsilon
    - \frac{
        \pi^2 \big(
            4 s^2 + 2 s^2 \log \!\big( \tfrac{1 + s^2}{4\pi^2} \big) + 3
        \big)
    }{
        4 \tilde{q}^2 (1 + s^2)^2
    } \varepsilon^2
    + \mathcal{O}(\varepsilon^3) \, , \\[1em]
    
\displaystyle
S_1 = \frac{\pi^2}{\tilde{q}^2 \sqrt{1 + s^2}}
    - \frac{2\pi^2}{\tilde{q}^2 (1 + s^2)^{3/2}} \, \varepsilon
    + \frac{
        \pi^2 \big(
            s^2 + 4 s^2 \log \!\big( \tfrac{1 + s^2}{4\pi^2} \big) + 6
        \big)
    }{
        2 \tilde{q}^2 (1 + s^2)^{5/2}
    } \varepsilon^2
    + \mathcal{O}(\varepsilon^3) \, , \\[1em]
    
\displaystyle
S_{\mathrm{log},0} =-
    \frac{\pi^2 s^2}{\tilde{q}^2 (1 + s^2)^2} \, \varepsilon^2
    + \mathcal{O}(\varepsilon^3) \, , \\[1em]
    
\displaystyle
S_{\mathrm{log},1} =
    \frac{4 \pi^2 s^2}{\tilde{q}^2 (1 + s^2)^{5/2}} \, \varepsilon^2
    + \mathcal{O}(\varepsilon^3) \, .
\end{cases}
\end{equation}
If we restrict our analysis to linear order in $\varepsilon$, we would conclude that the deformed SYK model at $n=1+\varepsilon$ has a zero-temperature entropy that depends on $s$, and that its leading correction is linear in temperature \cite{Anninos:2022qgy}. However, including the $\varepsilon^2$ corrections, one realises that there are logarithmic corrections, meaning this expansion cannot be trusted a very low temperatures.

 This can be explained by the following observation made in appendix \ref{sec: appendix on conformal perturbation theory}. As $n$ gets closer to 1, the entropy receives more and more contributions of the form $(\bj)^{2-2n},(\bj)^{4-4n}, \ldots$ as in \eqref{multiple stronger-than-linear terms}. In a naive perturbative expansion around $n=1$, each of these terms would have logarithmic divergences. For instance, for the first one, which is the anomalous term from \eqref{eq: entropy n<3/2} with the coefficient \eqref{eq: tilde S1}, we obtain
\begin{equation}
    \label{anomalous with 1+eps}
    \frac{\tilde{S}_1(n,s)}{(\bj)^{2n-2}} \Bigg|_{n=1+\varepsilon}= \frac{\pi ^2}{2 s^2}\varepsilon +  \left(-\frac{\pi ^2  (1+\log (2\pi s))}{s^2}-\frac{\pi^2 \log(\bj)}{s^2}\right)\varepsilon ^2+\mathcal{O}(\varepsilon^3) \, .
\end{equation}
As can be seen, the logarithmic divergences only appear at second order in the $\varepsilon$ expansion.

\subsection{Expanding around $n=2$}
Another case that can be solved perturbatively is $n=2+\varepsilon$ with $0< \varepsilon \ll 1$. Note that the results here can be trivially extended to $n=2-\varepsilon$. We expand
\begin{equation}
    \label{2 plus epsilon expansion}
    g(t)=g_{0}(t)+g_1(t)\varepsilon+\mathcal{O}(\varepsilon^2) \, ,
\end{equation}
where the coefficients solve
\begin{equation}
    \label{2-eps ODEs}
    \begin{cases}
        \partial_{t}^2 g_0(t) &= 2 (\bj)^2 \left(2 s^2 e^{\frac{g_0(t)}{2}}+e^{g_0(t)}\right) \, , \\
        \partial_{t}^2 g_1(t) &= 2 (\bj)^2 \left(e^{g_0(t)} g_1(t)-\frac{1}{2} s^2 e^{\frac{g_0(t)}{2}} \left(g_0(t)-2 g_1(t)-2\right)\right) \, , \\
    \end{cases}
\end{equation}
with vanishing boundary conditions at $t=0,1$ for each coefficient. The leading term $g_0(t)$ is given in \eqref{n=2 solution} while the expression for $g_1(t)$ is cumbersome, and so we only include the value at $t=1/2$ which is sufficient to find the entropy following \eqref{Entropy}. It is given by
\begin{equation}
    \label{n=2 eps order solution}
    g_1\!\left( \frac{1}{2} \right) = \left(g_0\!\left( \frac{1}{2} \right) - 4 \right) \left(\frac{e^{g_0 \left(\frac{1}{2} \right)/2}}{s^2}+2 \right)^{-1} \, .
\end{equation}
Interestingly, this is the precise value for which there is no contribution of order $\varepsilon$ to the large $q$ entropy,\footnote{We make here the distinction with a large-$\tilde{q}$ expansion, for which there is a trivial term of order $\varepsilon$.} so that
\begin{equation}
    \label{2+epsilon entropy}
    \frac{S}{N}\Bigg|_{n=2+\varepsilon} = \frac{S}{N}\Bigg|_{n=2}+\mathcal{O}(\varepsilon^2) \, .
\end{equation}
Assuming that the $\varepsilon \ll 1$ limit commutes with the low-temperature limit, this would imply that the derivative of the large $q$ entropy with respect to $n$ vanishes at $n=2$. However, the numerics seem to contradict this implication, which indicates that the limits probably do not commute.

\subsection{Other values of $n$}\label{sec: appendix other n}

In section \ref{sec: more analytically solvable cases}, we formulated a method of computing the low-temperature entropy, in the deep IR, for particular values of $n$. We present here the results for the linear-in-temperature terms $S_1(n,s)$ in the cases $n=3$ and $n=4$. The former is given by
\begin{equation}
    \label{Linear coef n=3}
    \begin{aligned}
        S_1(n=3,s) = \frac{\pi^2}{s}\left[ -\frac{3i}{4}  F_1\left(\frac{1}{2};\frac{1}{2},-\frac{1}{2};\frac{3}{2};3 i s,-3 i s\right)+\frac{3}{2} s F_1\left(\frac{3}{2};\frac{1}{2},\frac{3}{2};\frac{5}{2};3 i s,-3 i s\right) \right. \\ \left.
        +s \, _2F_1\left(\frac{1}{2},\frac{3}{4};\frac{7}{4};-9 s^2\right)+\frac{3i}{2} \, _2F_1\left(\frac{1}{4},\frac{1}{2};\frac{5}{4};-9 s^2\right)+\frac{(2-2 i) \sqrt{6} E(2)-9 i \sqrt{\frac{(1-3is)s}{1+3is}} }{12 \sqrt{s}} \right] \, ,
    \end{aligned}
\end{equation}
where $_2F_1$ is the Gauss hypergeometric function, $F_1$ is the Appell hypergeometric function, and $E$ is the complete elliptic integral of the second kind. While for $n=4$, the result is
\begin{equation}
\label{Linear coef n=4}
    \begin{aligned}
S_1(n=4,s) = &\frac{\pi^2}{8 \, 2^{2/3} \, s^{4/3}} \,  \Bigg[ 
4\sqrt{2}\,\sqrt{\,3 + i\sqrt{3} 
- \frac{6 \left(1 - 2\cdot 2^{1/3}s^{2/3} + 4\cdot 2^{2/3}s^{4/3}\right)}{1+16s^2}} \\
&\qquad \times \, 
F_1\!\left(\tfrac{1}{2}, \tfrac{3}{2}, \tfrac{1}{2}; \tfrac{3}{2}; \,
\frac{-1+i\sqrt{3}+4\cdot 2^{1/3}s^{2/3}}{2+4\cdot 2^{1/3}s^{2/3}}, \,
\tfrac{1}{2} + \frac{i\sqrt{3}\left(1-2\cdot 2^{1/3}s^{2/3}\right)}{2+4\cdot 2^{1/3}s^{2/3}}
\right) \\
&\quad + \frac{3^{1/4}}{3i+\sqrt{3}} \,  \Bigg(
-16(-1)^{7/12} \,
F\!\left(\arcsin\!\sqrt{\tfrac{\tfrac{2i}{-i+\sqrt{3}}+2\cdot 2^{1/3}s^{2/3}}{1+2\cdot 2^{1/3}s^{2/3}}}, \;
\frac{1-i\sqrt{3}}{2}\right) \\
&\qquad\qquad 
+ \sqrt{2(-i+\sqrt{3})} \Big( 
4i(i+\sqrt{3})\,K\!\left(\tfrac{1}{2}(1-i\sqrt{3})\right) \\
&\qquad\qquad\qquad\qquad
+ (3+i\sqrt{3}) \pi \, {}_2F_1\!\left(\tfrac{3}{2},\tfrac{3}{2};2;\tfrac{1}{2}(1-i\sqrt{3})\right)
\Big) \Bigg) \Bigg] \, ,
\end{aligned}
\end{equation}
where $F$ is the incomplete elliptic integral of the first kind, and $K$ is the complete elliptic integral of the first kind.

\section{Small \texorpdfstring{$s$}{s} expansion of 
  \texorpdfstring{$S_1(n,s)$}{S1(n,s)}}
  \label{sec: S1 at small s appendix}

\begin{figure}[H]
    \centering

    \subfigure[]{
        \includegraphics[width=0.48\textwidth]{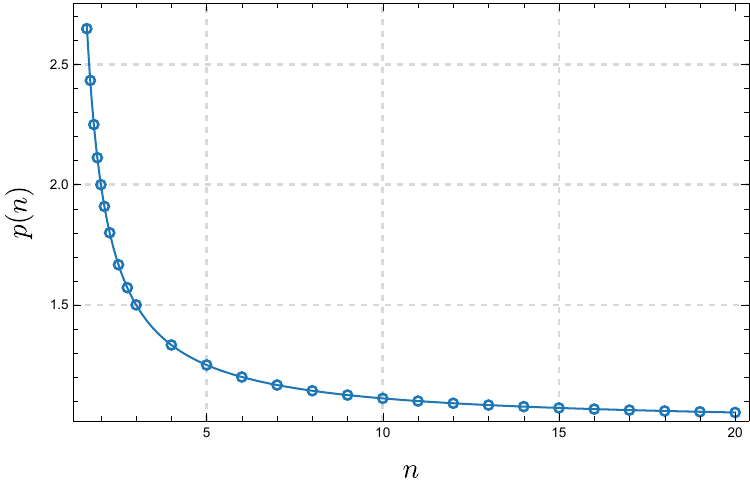}
    }
    \hfill
    \subfigure[]{
        \includegraphics[width=0.48\textwidth]{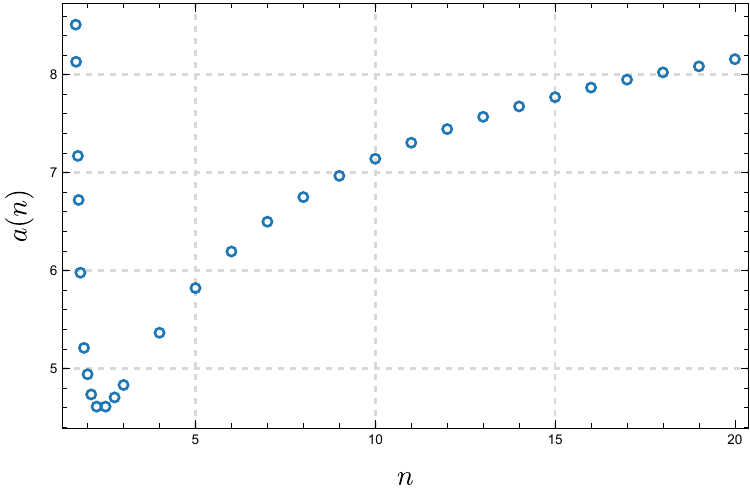}
    }

    \caption{The coefficients $p(n)$ and $a(n)$ from \eqref{small s S1} as a function of $n$. The circles are numerical evaluations while the solid curve in (a) corresponds to \eqref{p(n)}.}
    \label{fig: small s expansion}
\end{figure}

In the case $n > 3/2$, the entropy in the deep IR admits a dominant linear-in-temperature term whose coefficient $S_1(n,s)$ we were not able to compute analytically for a generic values of $n$. In this appendix, we show the numerical results for the expansion of $S_1(n,s)$ from \eqref{entropy n>3/2} at $s \ll 1$. We take values for $s$ of the order $10^{-6}$ and make a fit with the ansatz
\begin{equation}
    \label{small s S1}
    S_1(n,s \ll 1)=\frac{a(n)}{s^{p(n)}}+\cdots \, ,
\end{equation}
where $a(n)$ and $p(n)$ are to be determined. The results are given in figure \ref{fig: small s expansion}, where we also include a guess,
\begin{equation}
    \label{p(n)}
    p(n) = \frac{n}{n-1} \, ,
\end{equation}
which exactly matches the analytical results $n=2,3,4$ and seems to be an excellent fit for the numerical results.

\section{Deformed SYK model with multiple deformations} \label{app: multiple}

In this paper, we have identified the leading infrared thermodynamic behaviour of a class of deformations to the SYK model. Given these individual contributions, one may consider combinations thereof to construct more general thermodynamic behaviour. Such a combination is generally given by the model
\begin{equation}
    \label{mult deformations Hamiltonian}
    H=H_q+\sum_{\mathcal{N}} \, s(n)H_{q/n} \, \, ,
\end{equation}
for some non-negative function $s(n)$ supported on a set $\mathcal{N}$. Working in the large-$q$ limit, the equations of motion are given by
\begin{equation}
    \label{mult deformations EOM}
    \partial_{t}^2g(t)=2(\bj)^2 \left(e^{g(t)}+\sum_{\mathcal{N}} \, ns(n)^2e^{g(t)/n} \right) \,,
\end{equation}
supplemented by boundary conditions $g(0)=g(1)=0$. The entropy is now generalised to
\begin{equation}
\label{mult deformations Old Entropy}
    \frac{S}{N}=S_0^{\text{free}}+\frac{1}{8q^2} \int_{0}^{1}dt \left[\frac{1}{2}(\partial_tg(t))^2-2(\beta \mathcal{J})^2 \left(e^{g(t)}+\sum_{\mathcal{N}} \, n^2s(n)^2e^{g(t)/n}\right) \right] \,.
\end{equation}
As with the single deformation, the equations of motion simplify this integral into
\begin{equation}
\label{mult deformations Entropy}
    \frac{S}{N}=S_0^{\text{free}}-\frac{(\bj)^2}{4q^2}  \left(e^{g_m}+\sum_{\mathcal{N}} \, n^2s(n)^2e^{g_{m}/n}\right) \, ,
\end{equation}
where $g_m=g(t=1/2)$, which satisfies
\begin{equation}
    \label{bj in terms of gm mult def}
    \bj = \int_{g_m}^{0}\frac{dx}{\sqrt{W(x)-W(g_m)}} \, \, , \, \,
    W(x)=e^{x}+\sum_{\mathcal{N}} \, n^2s(n)^2e^{x/n} \, .
\end{equation}
Therefore, upon choosing a set $\mathcal{N}$ and a function $s(n)$ (to retain at least one intermediate IR regime, take $\sup_{n}s(n) \ll 1$), one can determine $g_m$ using \eqref{bj in terms of gm mult def}, plug it into \eqref{mult deformations Entropy} and evaluate the entropy along the thermal RG flow. This should be feasible numerically, and analytically in some very fine-tuned cases.

As an example, consider an SYK model with a very large but finite value of $q$. Then, there are many deformations with $\tilde{q}<q$ such that $1<q/\tilde{q}<3/2$, so that we may approximate the sum in \eqref{mult deformations Hamiltonian} by an integral. Then, motivated by the coefficient \eqref{eq: tilde S1} of the anomalous term in \eqref{eq: entropy n<3/2}, taking the interval $\mathcal{N}=(1.25,1.30)$ and
\begin{equation}
    \label{fine-tuned s(n)}
    s(n) =  \frac{1}{10}\left(\frac{\pi ^{2 n-\frac{1}{2}} \Gamma \left(\frac{1}{2}-n\right)}{2 n \Gamma (1-n)}-\frac{\pi ^{2 n-\frac{1}{2}} \Gamma \left(\frac{1}{2}-n\right)}{4 n^2 \Gamma (1-n)}\right)^{\frac{1}{2 n}} \, ,
\end{equation}
where the factor of $10$ is included to ensure $s(n) \ll 1$, we can expect the leading infrared behaviour of this multi-deformed model to be
\begin{equation}
    \label{multi-deformation example}
    \int_{1.25}^{1.30}dn \, \frac{\tilde{S_1}(n,s(n))}{(\bj)^{2n-2}} \propto \frac{1}{\sqrt{\bj}\log(\bj)} + \cdots \, ,
\end{equation}
where the dots are subleading at low temperatures. This is yet another thermodynamic contribution to the entropy that dominates the linear-in-temperature term.

In short, the particular deformations studied in this paper may serve as building blocks to construct particular low-temperature behaviour. Given the connection established with two-dimensional spacetimes in section \ref{flow geometrisation}, one can consider reversing the logic. Given a macroscopic two-dimensional spacetime, characterised by a dilaton potential $U(\phi)$, we may make use of these deformed SYK models to produce microscopic large-$N$, strongly coupled quantum mechanical systems with the same thermodynamic behaviour.

A natural question is which macroscopic dilaton potentials can actually arise from an underlying microscopic quantum-mechanical model. For example, any Hermitian Hamiltonian necessarily yields an entropy that increases monotonically with temperature. This constraint effectively rules out the possibility of a unitary microscopic system in a thermal state whose emergent dilaton potential would generate centaur geometries \cite{Anninos:2017hhn, Anninos:2018svg}, which flow from from $\mathrm{AdS}_2$ asymptotics to a $\mathrm{dS}_2$ interior.

If one nevertheless insists on exploring such dilaton potentials, one option is to relax hermiticity of the microscopic Hamiltonian. Indeed, it was shown that in the $n=2$ model, an analytic continuation of the coupling of the form $s = i \mathfrak{s}$ with $-1/2 < \mathfrak{s} < 1/2$ yields, at least at a formal level, a negative linear-in-temperature coefficient for the entropy \cite{Anninos:2020cwo, Anninos:2022qgy}. In the present work, we have extended this analysis by providing analytical solutions for the $n=3$ and $n=4$ models, both of which exhibit a linear-in-temperature entropy in the deep infrared. In each case, there exist complex values of $s$ for which the entropy remains linear while the coefficient becomes negative. The possibility of microscopically building a patch of de Sitter space in this context remains an interesting problem to explore in future work.

\bibliographystyle{JHEP}
\bibliography{bibliography}

\end{document}